\newcommand{\ignore}[1]{}

\documentclass[sigplan,screen]{acmart}
\AtBeginDocument{%
  }

\usepackage{physics}
\usepackage{algorithmic}
\usepackage{multirow}
\usepackage{enumitem}
\usepackage{tikz}
\usepackage[ruled]{algorithm2e}
\setcitestyle{comma}
\usepackage{listings}

\definecolor{aliceblue}{rgb}{0.94, 0.97, 1.0}
\usepackage{tcolorbox}
\tcbset{width=1\columnwidth,boxrule=1pt,colback=aliceblue,arc=1pt,left=2pt,right=2pt,boxsep=0.5pt}

\newtcolorbox{hintbox}[2][]
{
  colframe = oldaliceblue!100,
  colback  = oldaliceblue!5,
  boxsep=2pt,
  width=\dimexpr\columnwidth\relax, 
  coltitle = oldaliceblue!20!black,
  title    = #2,
  #1,
}

\copyrightyear{2027}
\acmYear{2027}
\setcopyright{cc}
\setcctype{by}
\acmConference[ASPLOS '27]{Proceedings of the 32nd ACM International Conference on Architectural Support for Programming Languages and Operating Systems, Volume 1}{April 11--15, 2027}{Heraklion, Greece}
\acmBooktitle{Proceedings of the 32nd ACM International Conference on Architectural Support for Programming Languages and Operating Systems, Volume 1 (ASPLOS '27), April 11--15, 2027, Heraklion, Greece}
\acmDOI{10.1145/3845814.3850031}
\acmISBN{979-8-4007-3012-2/2027/04}

\begin{document}


\title[Trident: Three-Level Memory Hierarchy for Yoked Surface Codes]{Trident: Efficient Load Verification for Yoked Surface Codes using a Three-Level Memory Hierarchy}

\author{Gunagya Singh Mamak}
\orcid{0009-0009-7218-1596}
\email{gmamak3@gatech.edu}
\affiliation{%
  \institution{Georgia Institute of Technology}
  \city{Atlanta}
  \state{GA}
  \country{USA}
}

\author{Suhas Vittal}
\orcid{0000-0003-0236-701X}
\email{suhaskvittal@gatech.edu}
\affiliation{%
  \institution{Georgia Institute of Technology}
  \city{Atlanta}
  \state{GA}
  \country{USA}
}

\author{Kabir Dubey}
\orcid{0000-0002-7621-9004}
\email{kabir@u.northwestern.edu}
\affiliation{%
  \institution{Northwestern University}
  \city{Evanston}
  \state{IL}
  \country{USA}
}

\author{Kaitlin N. Smith}
\orcid{0000-0002-1169-3696}
\email{kns@northwestern.edu}
\affiliation{%
  \institution{Northwestern University}
  \city{Evanston}
  \state{IL}
  \country{USA}
}

\author{Moinuddin K. Qureshi}
\orcid{0000-0002-1314-9096}
\email{moin@gatech.edu}
\affiliation{%
  \institution{Georgia Institute of Technology}
  \city{Atlanta}
  \state{GA}
  \country{USA}
}


\begin{abstract}

\textit{Fault-Tolerant Quantum Computers (FTQCs)} commonly use the \textit{surface code} to encode program qubits as error-corrected logical qubits. Unfortunately, the surface code has a poor encoding rate, so recent work has proposed \textit{Yoked Surface Codes (YSCs)}, which concatenate the surface code with a high-rate quantum code. The surface code qubits in YSCs are protected by the checks, or \textit{yokes}, of the high-rate code. YSCs have 1D and 2D variants, which are $1.8\times$
and $4.5\times$
 more efficient than the surface code, respectively.

Modern FTQC architectures use a two-level hierarchy: surface codes perform logical operations (compute) while 2D YSCs store idle program qubits (memory). To load a qubit from memory, the FTQC first performs \textit{retrieval}, which logically moves a program qubit from a 2D YSC into a surface code (8 cycles). However, retrieval is not fault-tolerant, so the FTQC must \textit{verify} the load by measuring all the yokes of the 2D YSC (178 cycles). As verification dominates memory access latency, eliminating it can provide a speedup of $5.7\times$.

Our goal is to reduce the verification overhead of 2D YSCs while retaining their storage efficiency. We exploit the fact that 1D YSCs, though less storage-efficient, incur negligible verification latency. We propose \textit{Trident}, which expands the two-level hierarchy into a three-level hierarchy by adding a small intermediate level containing a 1D YSC (L1). If most loads hit L1 rather than L2 (2D YSC tier), it can significantly reduce verification overhead. We present two variants: \textit{Trident-D} (demand-driven) and \textit{Trident-P} (prefetch-driven), which achieve speedups of $2.8\times$
and $4.2\times$, respectively, while increasing the physical qubit cost of memory by only 4.9\%.

\ignore{
    This paper studies memory system bottlenecks in \textit{Fault-Tolerant Quantum Computers (FTQCs)}
    using the surface code.
    Modern FTQCs using the surface code contain separate compute and memory subsystems, where the memory
    subsystem is implemented using \textit{Yoked Surface Codes (YSCs)}.
    YSCs concatenate the surface code with high-rate algebraic codes to reduce physical qubit footprint.
    YSCs come in two variants: 1D YSCs which are $2\times$ more efficient than the surface code, and 2D
    YSCs with are $3\times$ more efficient than the surface code.
    Of the two, 2D YSCs are preferable as a memory.
    Loading from a 2D YSC is rather fast and only takes 12 cycles.
    However, YSC loads are not fault-tolerant as the YSC memory cannot concurrently serve
    memory accesses and measure the parity checks required to protect its constituent surface
    codes.
    Thus, the load must be subsequently \textit{verified} by measuring the YSC memory's parity
    checks: this takes 178 cycles.
    Our studies find that verification causes a $5.6\times$ slowdown relative to an idealized system
    with no verification latency.

    To this end, we propose using small, 8-entry 1D YSC as a compiler-managed cache 
    to mitigate verification latency.
    Our insight is that while 1D YSCs are less storage efficient, their verification
    latency is 6 cycles, which is negligible.
    Hence, if the cache has decent hit rate, verification-related slowdowns will reduce.
    We find that introducing this cache improves performance by $2.7\times$.
    However, caching performs poorly on workloads where the working set is rather large due
    to thrashing.
    On such workloads, verification latency largely remains on the critical path.
    To mitigate these overheads, we propose PROMPTLY, a software prefetcher that
    moves program qubits into the 1D YSC cache well before their corresponding load
    to hide their verification latency.
    {\color{red}However, the number of inflight prefetches is limited by the capacity of the cache, thus
    limiting coverage.
    PROMPTLY balances both coverage and timeliness by precisely scheduling prefetches only
    for the loads whose verification can be wholly hidden.}
    PROMPTLY improves performance by $4.2\times$, thus providing significant improvements
    upon only caching.
}

\ignore{
    Realizing practical applications on noisy quantum hardware will require fault-tolerant quantum computing based on
    quantum error correction (QEC). The surface code is the most prominent QEC scheme due to its hardware-friendly
    planar design and mature fault-tolerant instruction set. Despite being well-suited for active computations, the
    code's low encoding density makes large-scale programs spatially expensive. Recent yoked surface codes improve this
    density without sacrificing planar connectivity by protecting surface codes with periodic logical parity checks
    called yokes, yielding new 1D and 2D yoked code blocks. The superior density of 2D yoked surface codes makes them
    natural candidates for dense memory in a compute-memory hierarchy, but the data flow implications and performance
    impacts of this hierarchy remain understudied.

    In this paper, we present the first instruction-level simulation of programs executing on a yoked surface-code
    architecture. We show that qubits loaded from 2D yoked memory are not ready for arbitrary computation until the
    source block's yokes are checked. Since the 2D block comprises many yokes measured serially, this readiness delay
    inflates memory latency to hundreds of cycles and can bottleneck execution, causing a $5.4\times$ slowdown. To
    address this problem, we introduce a three-level architecture that places a small 1D yoked buffer---which checks
    only two yokes---between compute and 2D memory. We then present PROMPTLY, a prefetcher that stages qubits from 2D
    into the 1D buffer before they are needed in compute, absorbing the readiness latency of 2D memory off the critical
    path. PROMPTLY balances prefetch timeliness against coverage, and achieves a \(4.1\times\) speedup over a baseline
    architecture with compute and 2D memory only. Furthermore, we show that using 1D and 2D blocks in memory reduces
    fragmentation relative to a 2D-only design, delivering these performance gains with negligible space overhead.
}
\end{abstract}




\maketitle

\begin{figure*}
    \centering
    \includegraphics[width=0.95\textwidth]{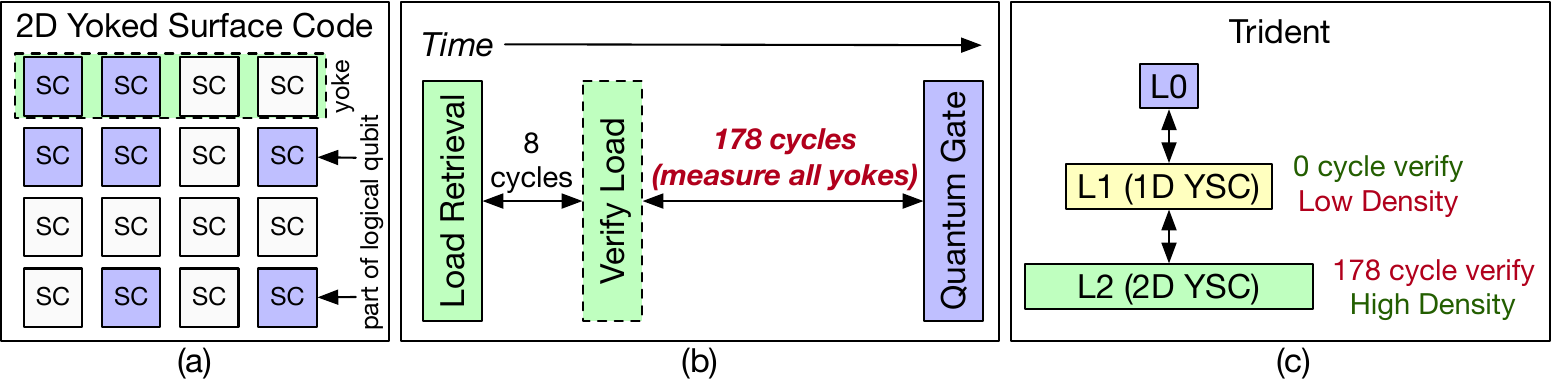}
    \caption{(\textbf{a})~2D yoked surface codes (YSCs) encode multiple surface code qubits and protect them via checks known as yokes.
             (\textbf{b})~Load operation from 2D YSCs.
                            Before the retrieved qubit can be used, we must verify that the load occurred successfully
                            by measuring the yokes, which incurs high latency (178 cycles).
             (\textbf{c)}~Our proposal, \textit{Trident}, is a three-level memory hierarchy which uses a single 1D YSC block (L1) to
                            reduce mean load verification latency.
             }
    \label{fig:intro}
\end{figure*}

\section{Introduction}
    Quantum computing provides exponential speedups for important
    applications, but the error rates required to execute these applications are typically $10^{-9}$ or less~\cite{shor1994_algorithms,
                                reiher2017_nitrogen_fixation,
                                kim2022_qchem_resource_estimates,
                                grover1996_algorithm}.
    \textit{Quantum Error Correction (QEC)} is 
    the most promising method of reaching such low error rates~\cite{calderbank1996_css_codes,
                    shor1995_qec}.
    QEC codes encode a large number of unreliable 
    \textit{physical qubits} into a \textit{code block} containing few \textit{logical qubits}. Throughout program execution, the parity checks of the QEC
    code are periodically measured to identify errors in the code block. The error correction capabilities of a QEC code can
    be increased in two ways.
    First, by increasing the 
    \textit{code distance} ($d$) by adding more
    qubits.
    Second, by \textit{code concatenation}, which
    uses the logical qubits of one QEC code 
    (\textit{inner code}) as
    the data qubits of another QEC code 
    (\textit{outer code}).


    In this paper, we study \textit{Fault-Tolerant Quantum Computers (FTQCs)}
    which use the surface code for error correction~\cite{kitaev1997_toric_codes,
                                                            dennis2001_surface_codes,
                                                            tomita2014_surface_code_with_realistic_noise}.
    Fault-tolerant computation involving the surface code is
    highly optimized~\cite{horsman2012_surface_code_lattice_surgery,
                            litinski2019_game_of_surface_codes,
                            litinski2019_msd,
                            gidney2024_cultivation,
                            gidney2024_inplace_y_basis,
                            fowler2018_lattice_surgery_computation}.
    Unfortunately, the surface code's encoding rate is poor,
    requiring nearly 1000 physical qubits per logical qubit in practice for $d=25$ codes. Recent work has proposed \textit{Yoked Surface Codes (YSCs)}
    to mitigate the overhead of surface codes~\cite{gidney2023_yoked_sc}.
    Figure~\ref{fig:intro}(a) shows an overview of
    YSCs.
    YSCs concatenate the surface code with
    high-rate quantum codes to reduce the physical qubit footprint.
    The parity checks of the high-rate quantum code, or \textit{yokes}, 
    protect the surface code qubits within the YSC.
    YSCs have two variants, 1D and 2D, which are $1.8\times$ and $4.5\times$
    more efficient than a normal surface code, respectively. Given the storage efficiency of 2D YSCs, modern FTQC studies
    advocate the use of 2D YSCs instead of 1D YSCs~\cite{babbush2026_ecc256,
                                            gidney2025_rsa_factorization,
                                            khattar2025_verifiable}. 


    While 2D YSCs have better storage efficiency than the normal
    surface code, they cannot both perform logical operations
    and measure yokes concurrently.
    Furthermore, stalling yoke measurements to execute
    logical operations on the YSC would increase the probability
    of a logical error.
    Therefore, modern FTQCs use a two-level \textit{exclusive} 
    memory hierarchy.
    The compute subsystem of the hierarchy uses a few
    (normal) surface codes to execute logical operations, whereas the memory subsystem uses the 2D YSC blocks to store the idle program qubits in a storage-efficient manner. Because the design devotes most of the physical qubits to the memory subsystem, the overall storage efficiency of the two-level hierarchy remains close to that of the 2D YSC. 

\subsubsection*{Accessing the Yoked Surface Code Memory:}
        Throughout program execution, the FTQC will move program
        qubits between the two subsystems using load and store
        operations.
        Loads and stores involve interacting with the constituent
        surface qubits within a 2D YSC block.
        These interactions consume the routing channel within the 
        2D YSC block which is normally used to measure the yokes
        of the block.
        At any given time, the channel can be used to measure
        yokes or serve memory accesses, but not both concurrently \cite{gidney2023_yoked_sc}.

        Unfortunately, this limitation means that neither loads
        nor stores are fault-tolerant~\cite{gidney_stack_exchange_yoke_blocking}.
        During a memory access, the surface codes constituting a
        2D YSC are unprotected by its yokes.
        These surface codes may incur logical errors that
 compromise the outcome of the load or store instruction.
        Hence, a loaded or stored program qubit is in an
        unverified state, and the FTQC cannot execute
        any logical operations until the retrieved logical qubit is verified.
    
\subsubsection*{The Verification Bottleneck of 2D YSCs:}
        Resolving the state of the program qubit is called
        \textit{verification} and requires the FTQC to measure
        all yokes of the 2D YSC block.
        For a 2D YSC block, verification incurs significant latency overheads.
        For a store, this verification latency is off the critical
        path as compute instructions do not depend on stores.
        In contrast, compute instructions are often dependent
        on the load.
        So, until load verification completes, the FTQC must stall.
        Figure~\ref{fig:intro}(b) describes the operation of a load instruction, including retrieval (moving from the memory system to the compute system) and verification (ensuring that the retrieved qubit is free of error).
        While the retrieval of the program qubit only
        lasts eight cycles, verification lasts 178 cycles.
        If we could eliminate all verification-induced stalls (ideal), we would speed up execution by $5.7\times$, on average.
        
        Verification latency scales with the number of yokes
        in a 2D YSC as all yokes must be measured to verify
        a load.
        A 194-entry 2D YSC, for example, has 62 yokes which
        take 178 cycles to measure.
        In contrast, 1D YSCs only have two yokes, and these
        yokes can be interleaved with a memory access.
        Hence, the entire load operation, retrieval and
        verification, takes eight cycles, and the exposed
        latency of verification is \textit{zero}.
        Unfortunately, while replacing all 2D YSC blocks with
        1D YSC blocks would reduce load latency, the physical qubit
        footprint of the memory subsystem would increase by 
        $2.5\times$.
        \textit{The goal of this paper is to reduce the
         verification latency of 2D YSCs while retaining
        their storage efficiency.}

\subsubsection*{A Case for a Three-Level Memory Hierarchy:}
        We propose \textit{Trident}, which expands the two-level
        memory hierarchy into a three-level memory hierarchy by including an intermediate memory tier consisting of a small 1D YSC (e.g., with eight entries). An overview of Trident is shown in Figure~\ref{fig:intro}(c). 
        We refer to the compute subsystem, 1D YSC memory,
        and 2D YSC memory of Trident as the \textit{L0}, \textit{L1}, and \textit{L2}, respectively.
        Our insight is that if a majority of loads in Trident access the
        L1 instead of the L2, then we can obtain a significantly reduced load verification
        latency, on average.

  \subsubsection*{Demand-Driven Hierarchy:}
        We first study \textit{Trident-D}, which operates as a 
        demand-driven hierarchy where the L1 operates as a cache.
        While this cache can be operated at runtime using the popular 
        LRU replacement policy, we observe that quantum programs have deterministic
        access patterns already known at compile-time, so we can obtain optimal (OPT) replacement.
        Trident-D implements the L1 as a compiler-managed cache.
        When a program qubit is to be stored into the L1, Trident-D
        must allocate space for the incoming program qubit.
        So, Trident-D evicts the qubit whose next load is farthest into
        the memory access stream and places the evicted qubit into
        the L2 tier.

    \subsubsection*{Prefetch-Driven Hierarchy:}
        The performance of a cache is limited by the temporal locality
        of the memory access stream.
        Ideally, we want a solution whose speedups are not limited by
        the access pattern of the program.
        To this end, we study \textit{Trident-P} which operates as a
        prefetch-driven hierarchy.
        For each load in the quantum program, Trident-P checks if the
        target program qubit resides in the L1 or L2.
        If it resides in the L2, Trident-P will prefetch the program
        qubit into the L1, well in advance of the intended use.
        To allocate space for this qubit, Trident-P will evict the oldest
        resident of the L1.
    
        Our studies find that Trident-D and Trident-P provide speedups of $2.8\times$ and
        $4.2\times$ relative to the baseline two-level memory hierarchy.
        Trident increases the physical qubit footprint of memory storage by only 4.9\% on
        average.

\vspace{0.05in}\noindent
    Overall, this paper makes the following contributions: 

\begin{figure*}[!htb]
    \centering
    \includegraphics[width=\textwidth]{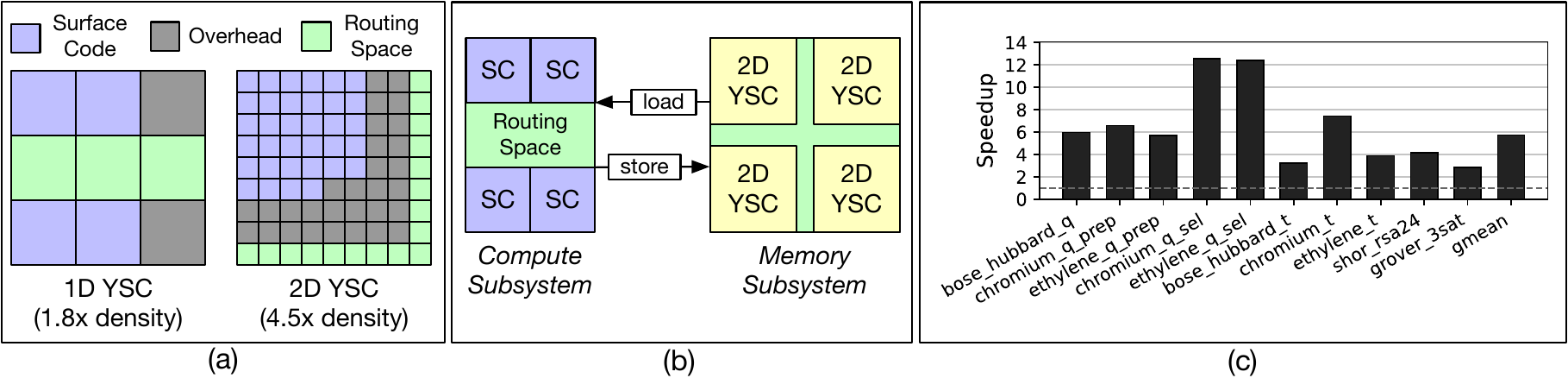}
    \caption{(\textbf{a})~Overview of 1D and 2D YSCs.
                            1D YSCs encode $n$ surface code qubits into $n-2$ logical qubits.
                            2D YSCs encode $n^2$ surface code qubits into $n^2 - 4n + 2$ logical qubits.
                            Both variants require additional routing space to measure yokes.
             (\textbf{b})~Overview of the two-level hierarchy used in modern FTQCs that use 2D yoked
                            surface codes.
             (\textbf{c})~Speedup of an idealized FTQC with zero verification latency.
                            As shown, the idealized FTQC achieves a gmean $5.7\times$ speedup.
                            }
    \label{fig:background}
\end{figure*}

        \begin{enumerate}
        [leftmargin=0.15cm,itemindent=0.5cm,labelwidth=\itemindent,labelsep=0.1 cm, align=left, itemsep=0.1cm, listparindent=0.5cm, topsep=0.15cm]
            \item To the best of our knowledge, this is the first paper to study the
                     performance impact of verification for 2D YSC.
            \item We propose \textit{Trident}, which
                    expands the FTQC's two-level memory hierarchy
                    into a three-level hierarchy which uses
                    a single 1D YSC as an intermediate memory tier (L1).
            \item We propose \textit{Trident-D} and \textit{Trident-P}
                     as demand-driven (L1 as cache) and 
                    prefetch-driven (L1 as prefetch buffer) hierarchies, respectively. They provide speedups of $2.8\times$ and
        $4.2\times$, respectively, over a two-level hierarchy. 
        \end{enumerate}


\ignore{
    Quantum computing has the potential to speed up several important but classically intractable problems such as integer
    factorization, quantum chemistry simulation, and unstructured search \cite{shor1994algorithms, reiher2017elucidating,
    kim2022fault, grover1996fast}. Running these algorithms at practically useful scales requires billions to trillions of
    operations \cite{gidney2025factor}. Since quantum hardware is too noisy to sustain such long computations, quantum error
    correction (QEC) is needed to suppress error rates \cite{calderbank1996good}. QEC encodes a smaller number of reliable
    logical qubits into many noisy physical qubits and continuously detects and corrects errors to preserve quantum
    information. Fault-tolerant quantum computers (FTQC) execute programs on these logical qubits using logical operations
    supported by the underlying error-correcting code.

    Among QEC schemes, the surface code has been the central focus in recent years due to its strong error resilience,
    compatibility with planar hardware, and highly optimized methods for logical computation \cite{fowler2012surface,
    google2025quantum, litinski2019game}.  Surface codes encode logical qubits as patches of a two-dimensional grid of
    physical qubits and repeatedly measure parity checks between neighboring qubits. It supports a well-developed
    fault-tolerant instruction set based on lattice surgery \cite{fowler2018low}. Thus, many proposed quantum computer
    architectures use the surface code as their computational substrate \cite{khattar2025verifiable, sharma2026space,
    stein2025_hetec}. Despite these advantages, the surface code is spatially expensive: a distance-$d$ patch requires
    $2(d+1)^2$ physical qubits. Large-scale applications are expected to require code distances on the order of $d=25$,
    meaning that a single logical qubit would occupy $2(25+1)^2=1352$ physical qubits. Since applications may themselves
    require thousands of logical qubits, the total hardware required can easily exceed one million physical qubits
    \cite{gidney2021factor}. Building quantum hardware at this scale is extremely challenging, motivating denser schemes for
    storing logical information.

    The surface code’s suitability for computation but inefficiency for storage motivate segregated compute--memory
    hierarchies. One way to obtain denser storage is to use alternative QEC codes, such as bivariate bicycle codes
    \cite{yoder2025_tour_de_gross}. However, such codes require non-local interactions between physical qubits, demanding
    newer hardware technologies that are still in their infancy \cite{bravyi2024high}. Ideally, we would want a denser
    encoding that remains compatible with existing planar hardware. This design goal forms the basis for recent work on
    yoked surface codes, which layers additional logical parity checks across groups of surface code patches
    \cite{gidney2025yoked}. These additional checks, called \textit{yokes}, can be measured using lattice surgery
    constructions that preserve planar connectivity. By adding protection across groups of patches, yokes improve
    reliability without paying the quadratic overhead of increasing the size of every patch, enabling up to $3\times$ higher
    storage density at relevant error rates. Thus, yoked surface codes are natural candidates for dense backing memory on
    planar devices.

    Two families of yoked surface codes are known: 1D blocks, which protect a linear array of patches with two yokes, and 2D
    blocks, which protects a grid array of patches with many row- and column-like yokes. Of the two, 2D blocks provide
    higher storage density, making them attractive as backing memory in surface code architectures \cite{gidney2025factor,
    khattar2025verifiable}. To maintain their target error rate, yoked blocks must periodically measure all of their yokes
    to remove patch-level logical errors. Accordingly, during program execution, a yoked block alternates between servicing
    memory accesses and performing a \textit{yoke-check sequence}; a memory operation cannot coincide with a yoke check.
    This creates a direct contention between data movement and periodic checks needed to preserve their logical fidelity,
    especially in large 2D blocks whose yoke check sequences span a significant interval of time. Hence, moving qubits
    between dense yoked storage and an active compute region introduces non-trivial latency and data flow constraints whose
    performance consequences remain understudied.

    When a qubit stored in a 2D block is needed for computation, it can be transferred into the compute region using
    lattice-surgery-based load operations. However, during this load, the qubit is not fully protected by the 2D block's
    yokes, since these cannot be checked simultaneously with the transfer. The probability of a logical error during this
    operation is consequently higher, which can result in an erroneous state being loaded into the compute region. This
    fault remains detectable, but it is not resolved until the source block completes its next yoke-check sequence. Until
    the loaded state is fully resolved, the qubit cannot safely participate in subsequent computation. We refer to the delay
    between load completion and resolution of the loaded state as \textit{readiness latency}. A 2D block consists of many
    yokes which must be checked serially, so one full yoke-check sequence can span hundreds of logical cycles. So, a qubit
    loaded from 2D memory may remain unavailable for a long time. When a pending operation on this qubit lies on the
    critical path, this readiness latency directly bottlenecks execution. Our simulations show that this occurs frequently
    in practice, resulting in a $5.8\times$ slowdown in program throughput relative to an ideal memory system.

    In contrast, a 1D block contains only two yokes, both of which can be checked in just a few cycles. Thus, loads from a
    1D block incur substantially lower readiness latency than loads from a 2D block. However, 1D storage carries roughly
    50\% higher spatial overhead than 2D storage, making it too expensive to serve as the sole backing memory. At the same
    time, it remains twice as dense as unyoked storage. This places 1D blocks in a useful middle ground between unyoked
    patches and 2D blocks in both density and access latency. We exploit this property to design a three-level architecture
    consisting of an unyoked compute region (0D), a small intermediate buffer built from 1D blocks, and a large main memory
    built from 2D blocks.

    A key challenge is managing data placement in the 1D buffer: which qubits should be placed there, when they should be
    placed, and which residents should be evicted when space is needed. Given the buffer's limited capacity and lower access
    latency, a natural first approach is to use it as a cache. Under this policy, a qubit is initially loaded from 2D memory
    into compute, operated on, and then written back into the 1D buffer. Once the buffer is full, standard replacement
    policies such as least-recently used (LRU) or Belady's optimal policy can be used to evict the resident qubit with the
    lowest anticipated utility back to 2D memory. These policies exploit temporal locality: repeated accesses to buffered
    qubits can be serviced from 1D rather than from 2D, yielding a \(2.8\times\) improvement in instruction throughput over
    a baseline architecture with only 0D compute and 2D memory. However, cache-based placement is fundamentally reactive,
    since a qubit enters the 1D buffer only after it has already been fetched from 2D into compute. If successive memory
    accesses are spread across many distinct qubits, the policy performs poorly on first accesses and leaves a substantial
    gap to ideal-memory performance.

    To overcome this limitation, we use the 1D region not just as a cache, but as a latency-hiding buffer. We propose
    \textbf{Pro}active \textbf{M}emory \textbf{P}refetching for a \textbf{T}hree-\textbf{L}evel \textbf{Y}oked Surface Code
    Architecture (\textit{PROMPTLY}). \textit{PROMPTLY} proactively stages qubits from 2D memory into the 1D buffer before
    they are needed in compute. This allows the 2D block to complete its long yoke-check sequence while the qubit resides in
    1D storage, absorbing most of the 2D readiness latency off the critical path of computation. When the qubit is
    ultimately needed in compute, it is loaded from the buffer and incurs only the short readiness latency of the 1D block.
    As with cache-based placement, a prefetched qubit must evict a resident from the 1D buffer to make room, and this
    eviction can still be guided by standard replacement policies. In this sense, prefetching subsumes cache-based
    replacement while providing additional coverage over first accesses that a cache cannot address.

    This latency-hiding depends critically on the \textit{timeliness} of the prefetch --- it must be issued early enough to
    hide a significant portion of the 2D yoke-check sequence. Since the program's memory accesses are known statically,
    software-inserted prefetches are highly accurate and can achieve high coverage. However, a prefetch issued immediately
    before the corresponding load is not sufficiently timely, since only a small fraction of the 2D readiness latency can be
    absorbed before the qubit is brought into compute, where the remaining delay stalls execution. At the other extreme,
    prefetching a qubit whose load lies far in the future improves timeliness but reduces effective coverage, since the
    qubit may occupy the limited 1D buffer long before it is needed, blocking other useful prefetches. \textit{PROMPTLY}
    addresses this tradeoff with a latency-aware policy that inserts prefetches only when they occur at a sufficient
    distance from the corresponding memory operation. Across our benchmarks, this approach outperforms reactive cache-based
    placement and reduces the performance gap to ideal memory, yielding a \(3.6\times\) speedup over baseline architectures.

    Finally, we study fragmentation in yoked surface code memory systems. Although 1D blocks are less dense than 2D blocks,
    counter-intuitively, a 1D+2D memory can occasionally improve density over a 2D-only design. This arises from the
    discrete-sizing constraints of 2D blocks, which can introduce internal fragmentation by forcing the system to provision
    more space than is needed in order to store a target number of qubits. By offering finer-grained sizing, 1D blocks can
    reduce this wasted space. We quantify this effect across a range of memory capacities. Accounting for fragmentation, we
    find that adding a 1D buffer introduces negligible ($<1\%$) space overhead while delivering meaningful performance
    gains.

    In summary, this work makes the following contributions:
    \begin{itemize}
        \item We present the first instruction-level simulation framework for programs executing on yoked surface code
        architectures.
        \item We identify and quantify the impact of readiness-latency of loads from a 2D yoked surface code memory and show
        that it can bottleneck existing systems.
        \item We introduce a three-level architecture consisting of 0D compute, 1D buffer and 2D main memory to balance
        access latency with storage density.
        \item We present \textit{PROMPTLY}, a proactive prefetcher that stages qubits through the 1D buffer to hide 2D
        readiness latency and improve performance beyond reactive cache-based placement.
        \item We show that combining 1D and 2D blocks can reduce fragmentation with negligible space overhead.
    \end{itemize}
}

\section{Background}

\subsection{Surface Codes}
    The surface code is a quantum error correcting code that encodes $2(d+1)^2$ physical
    qubits into a single, distance $d$ logical qubit~\cite{gidney2025_rsa_factorization,
                                                            babbush2026_ecc256, 
                                                            khattar2025_verifiable}.
    $d^2$ of these physical qubits are \textit{data qubits} that encode the logical state of
    the surface code.
    The remaining qubits are known as \textit{parity qubits}.
    During program execution, the data qubits will inevitably accumulate errors which modify
    the logical state of the surface code.
    Detecting these errors requires periodically performing \textit{syndrome extraction},
    which involves executing a physical circuit that entangles the data qubits 
    with the parity qubits.
    Then, the parity qubits are measured to obtain a syndrome that is analyzed by a classical
    \textit{decoder} to identify changes to the logical state.
    Any changes manifest as $X$ or $Z$ flips on the logical state and are tracked in software.

    Note that the syndrome extraction circuitry is also error-prone, and so the decoder cannot
    reliably identify changes to the logical state using only a single syndrome.
    Instead, the decoder must analyze $d$ consecutive rounds of syndromes to accurately
    filter time-like errors introduced by readout or $CX$ gates.
    Thus, $d$ syndrome extraction rounds constitute an \textit{error correction cycle}
    of the surface code.

\subsection{Fault-Tolerant Quantum Computation}
    \textit{Fault-Tolerant Quantum Computers (FTQCs)} use quantum error correction to protect program information
    from errors that occur during program execution.
    The surface code is a common choice for fault-tolerant quantum computation given that
    surface code computation is highly optimized~\cite{litinski2019_game_of_surface_codes,
                                                        litinski2019_msd,
                                                        gidney2024_cultivation,
                                                        gidney2024_inplace_y_basis,
                                                        horsman2012_surface_code_lattice_surgery,
                                                        fowler2018_lattice_surgery_computation}.
    Surface code logical operations are implemented using the \textit{lattice surgery}
    framework.
    Lattice surgery involves manipulating the boundaries of surface code logical qubits
    to measure Pauli-product operators, such as $ZZ$, between two logical qubits.
\ignore{
    The boundaries of the surface code correspond to its $X$ and $Z$ operators.
    In this paper, we take the convention that the left/right boundaries of the surface code
    corresponds to the $X$ operators, and the top/bottom boundaries correspond to
    the $Z$ operators, as depicted in Figure~\ref{fig:lattice_surgery}(a).
}
    Measuring a Pauli-product operator involves measuring the parity checks connecting the
    boundaries of two logical qubits for an entire error correction cycle.
    At the end of the error correction cycle, the decoder analyzes the $d$ rounds of syndrome
    data to resolve the measurement outcome.
    Logical operations, such as $CX$ gates, involve one or more 
    Pauli-product measurements~\cite{fowler2018_lattice_surgery_computation}.
    The reliable execution of a logical operation on a logical qubit is 
    contingent on the entire syndrome history of that qubit being available.
    Otherwise, the FTQC must stall until the history is available (see Appendix~\ref{sec:app:measurement_resolution}).

\ignore{
    \begin{figure}[!t]
        \centering
        \includegraphics[width=0.7\columnwidth]{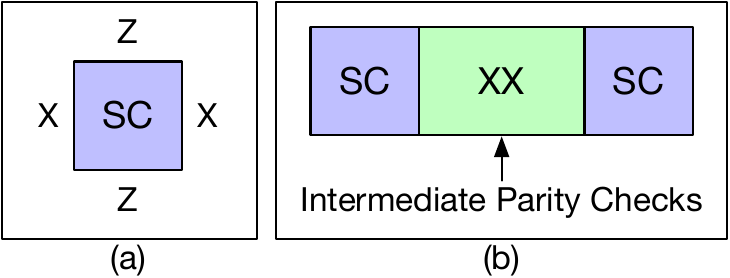}
        \caption{(\textbf{a})~The boundaries of a surface code.
                        The top/bottom boundaries are the $Z$ operator of
                        the logical qubit, and the left/right boundaries
                        are the $X$ operator.
                (\textbf{b})~The $X$ operators of two logical qubits can be
                    connected via lattice surgery to perform an $XX$
                    measurement.}
        \label{fig:lattice_surgery}
    \end{figure}
}

\ignore{
\begin{figure*}[!htb]
    \centering
    \includegraphics[width=0.7\textwidth]{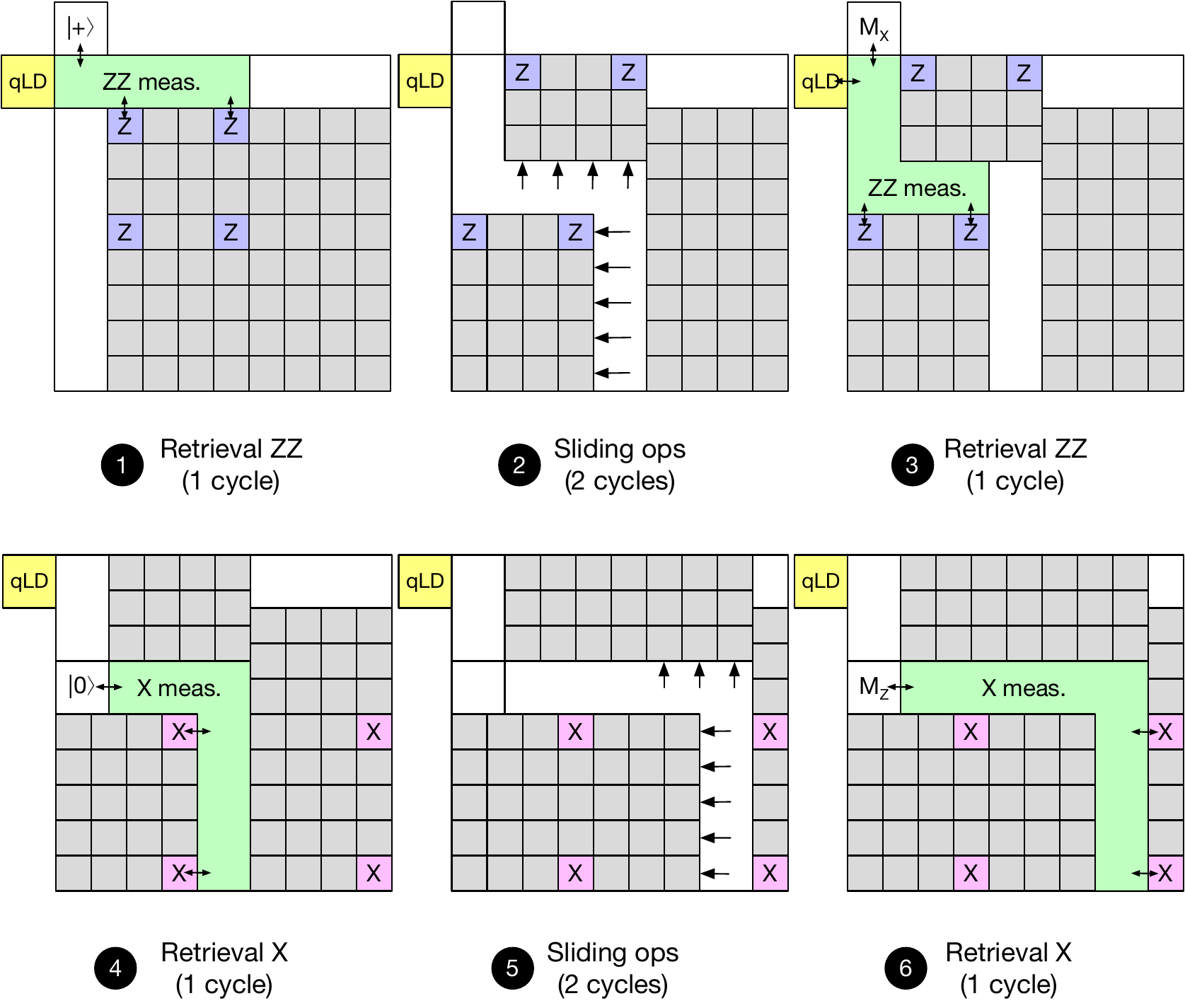}
    \caption{The series of lattice surgery operations required to load
            a program qubit from a 2D YSC block.
            The overall latency of a load is eight cycles.}
    \label{fig:2d_ysc_load}
\end{figure*}
}

\subsection{Yoked Surface Codes}
    The primary limitation of the surface code is its poor encoding rate, which reduces quadratically with the
    code distance ($d$).
    For low logical error rates (i.e., $10^{-15}$), a surface code logical qubit will require around $d = 25$ which
    consumes 1.3k physical qubits.
    To address this, recent works have proposed \textit{Yoked Surface Codes (YSCs)}~\cite{gidney2023_yoked_sc}, which
     concatenate the surface code with high-rate quantum codes with low code
    distance.
    The parity checks of the outer codes, or the \textit{yokes}, protect the inner surface code logical
    qubits.
    The effective distance of YSC ($d_\text{eff}$) scales with the product of the inner surface code
    distance ($d_\text{in}$) and the outer code distance.

    There are two variants of YSCs: 1D and 2D.
    Figure~\ref{fig:background}(a) presents an overview of both.
    1D YSCs concatenate $d_\text{in} = 15$ surface codes with an $[[n, n-2, 2]]$ quantum code and are $1.8\times$
    more efficient than normal surface codes.
    They have two yokes, one $X$ and one $Z$, measured across all $n$ surface code qubits, much like
    a checksum.
    2D YSCs concatenate $d_\text{in} = 11$ surface codes with an $[[n^2, n^2 - 4n + 2, 4]]$ quantum code and are
    $4.5\times$ more efficient than normal surface codes.
    The $n^2$ surface code qubits are organized into an $n \times n$ grid, and each yoke spans 
    one or two rows/columns of the grid.

\subsection{Two-Level Memory Hierarchy using 2D YSCs}
\ignore{
    Each logical qubit within a YSC corresponds to multiple constituent surface code logical qubits.
    Consequently, the $X$/$Z$ logical operator of each YSC logical qubit corresponds to
    the product of the $X$/$Z$ logical operators of the constituent surface code qubits.
    In principle, YSC blocks can perform quantum computation using lattice 
    surgery~\cite{gidney2023_yoked_sc}.
    Unfortunately, YSCs cannot concurrently perform computation and measure yokes
    as both require routing space to perform lattice surgery.
    While the FTQC can delay yoke measurements to perform computation, this would
    severely degrade the logical error rate.
    So, YSC-based computation is suboptimal.
}
    Modern surface code FTQCs use YSCs as a large, high-density memory for storing
    idle program qubits during program execution~\cite{gidney2025_rsa_factorization, 
                                                        babbush2026_ecc256, 
                                                        khattar2025_verifiable}.
    Figure~\ref{fig:background}(b) presents a high-level overview of modern
    FTQC architectures that use a \textit{two-level memory hierarchy}.
    The compute subsystem contains a few surface code logical qubits which execute
    logical operations such as $CX$ or $T$ gates.
    The memory subsystem uses one or more 2D YSCs for storage.
    Note that due to the no cloning theorem, the memory hierarchy is \textit{exclusive}.
    So, a program qubit can reside at only one level in the memory hierarchy.

\subsection{Loading from a Yoked Surface Code Memory}
    Throughout program execution, the FTQC will execute \textit{load} and \textit{store}
    instructions to move program qubits between the compute and memory subsystems. As the implementations of loads and stores are similar, we focus on loads for brevity. 
    
    In a YSC memory, a load has two parts: 
    \emph{retrieval} and \emph{verification}. 
    Accessing a YSC memory is much like accessing the DRAM-based
    memory system in traditional processors that use error-correcting codes (ECC) to correct
    data loaded from memory.
    In traditional systems, verification, where ECC is leveraged to
    detect or correct errors on the loaded data, incurs negligible
    overhead.
    Retrieval dominates load time.
    In contrast, in a YSC memory, we find that verification constitutes
    the majority of the load latency.
    Below, we discuss the technical details of retrieval and verification
    in YSCs.

    \subsubsection{Load Retrieval}
        In general, load retrieval from a YSC-encoded qubit \texttt{qA} into a surface
        code qubit \texttt{qB}
        (prepared in the $\ket{+}$ state) involves a logical-level state teleportation performed in two steps~\cite{gidney2023_yoked_sc, cohen2022_ldst}.
        First, the FTQC measures the joint $ZZ$ operator between \texttt{qA} and \texttt{qB} using
        lattice surgery.
        Then, \texttt{qA} is destroyed by measuring it in the $X$ basis.
        \texttt{qB} is left in the original state of \texttt{qA}.
        The two steps of load retrieval are complicated by the YSC encoding of \texttt{qA}.
        The $X$ and $Z$ logical operators of \texttt{qA} span multiple
         surface code patches within the YSC.
        Measuring the joint $ZZ$ operator therefore
        requires lattice surgery involving \texttt{qB} and all patches supporting the logical $Z$ operator of \texttt{qA}.
        Similarly, measuring \texttt{qA} in the $X$ basis requires 
        lattice surgery involving all patches supporting its logical $X$ operator, as measuring these patches transversally would destroy the \textit{entire} memory, not just \texttt{qA}.

    \subsubsection{Load Verification}
        During load retrieval, a YSC cannot measure its yokes since the routing space is used for
lattice surgery operations. If any of the surface code qubits
involved in these lattice surgery operations incur logical errors, then the outcome of the load is erroneous. Thus, loading
from a YSC memory is not fault-tolerant \cite{gidney_stack_exchange_yoke_blocking}. Recovering
from any errors encountered during the load operation requires the FTQC to measure all yokes of the YSC after the
load has finished. We refer to this process as verification.

Until the load is verified, the syndrome history of the
loaded program qubit is incomplete. Consequently, if a lattice surgery operation were to be executed using the loaded
program qubit, the decoder would be unable to accurately
resolve the operation’s outcome. Hence, the FTQC cannot
execute any gates using the loaded program qubit until verification completes the qubit’s syndrome history.
     
    \subsubsection{1D YSC: Retrieval and Verification}
\begin{figure}[!htb]
            \centering
            \includegraphics[width=1\linewidth]{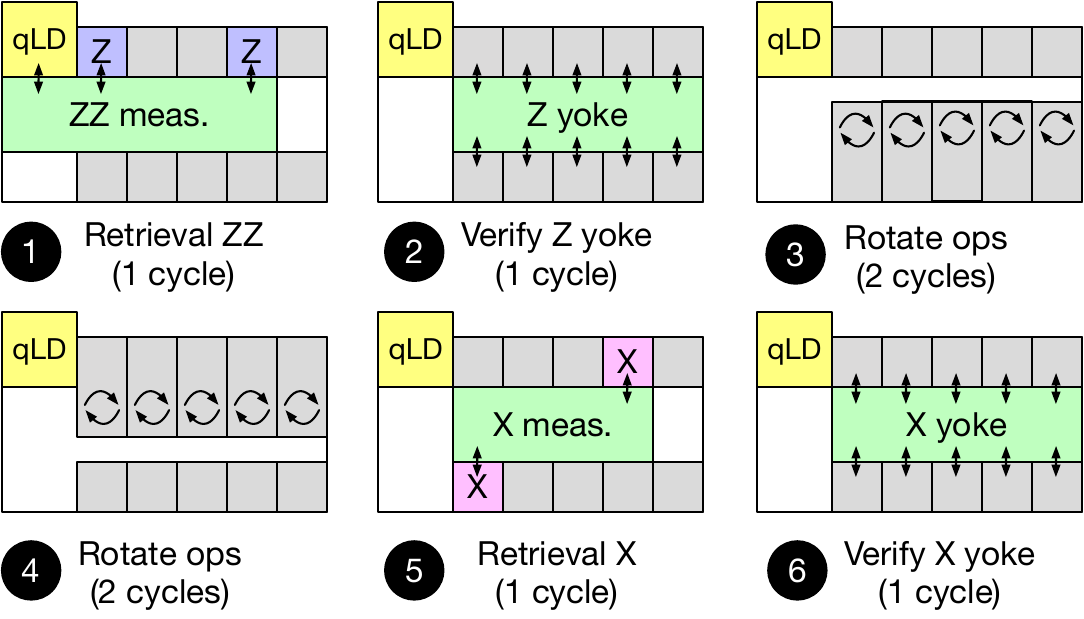}
            \caption{Load verification in 1D YSC. Yoke checks can be interleaved (steps 2 and 6) while loading from a 1D YSC.}
            \label{fig:1d_load}
        \end{figure}

        As the 1D YSC has only two yokes,
     retrieval and verification steps can be interleaved.
        Figure~\ref{fig:1d_load} shows the complete load procedure of
        a program qubit within the 1D YSC into a surface code qubit
        \texttt{qLD} initialized in the $\ket{+}$ state.
        The \textit{first retrieval step} measures 
        the joint $ZZ$ operator between \texttt{qLD} and the 
        target program qubit.
        As the program qubit is encoded using a 1D YSC, \texttt{qLD}
        performs $ZZ$ measurements with each surface code
        qubit constituting the program qubit's $Z$ operator.
        Then, the \textit{first verification step} measures
        the $Z$ yoke of the 1D YSC.
        As retrieval and verification both must access the $X$ operators
        of the 1D YSC's surface code qubits,
        the boundaries of each surface code qubit are ``rotated'', which takes four cycles in total.
        Next, the \textit{second retrieval step}
        measures the $X$ operator of the program qubit.
        This destroys the program qubit, leaving \texttt{qLD} in
        the program qubit's original state.
        The \textit{second verification step}
        measures the $X$ yoke of the 1D YSC.
        Hence, for a 1D YSC, a full load operation takes eight cycles.

    \subsubsection{2D YSC: Retrieval and Verification}
        Appendix~\ref{appendix:2d_load} discusses the specific lattice
        surgery operations required to implement load retrieval
        within 2D YSCs.
        Retrieval takes eight computation cycles. However, retrieval occupies the routing space required for yoke measurements, so the retrieved qubit remains unverified.
        
        For verification, yokes across all rows and columns are measured through a long sequence of lattice surgery operations. Specifically, measuring all yokes in 2D YSCs with width $w$ and inner surface-code distance $d_{\mathrm{in}}$ requires
        $25d_{\mathrm{in}}w + 4d_{\mathrm{in}}$
        rounds of syndrome-extraction \cite{gidney2023_yoked_sc}. For example, 2D YSCs with outer code $[[256,194,4]]$ ($w=16$) and inner distance $d_{\mathrm{in}}=11$ take $4444$ rounds of syndrome-extraction for verification. For this code, one computation cycle comprises $d_{\mathrm{eff}}=25$ rounds, so verification takes $4444/25 \approx 178$ cycles. Thus, following retrieval, a loaded program qubit must
        wait 178 cycles for its load to be verified.

\subsection{Motivation and Goal}
    Modern FTQCs that use the surface code use 2D YSCs within the memory subsystem
    of their two-level memory hierarchy.
    While the load retrieval latency of 2D YSCs is rather short (eight cycles), their
    load verification latency is lengthy (178 cycles).
    Figure~\ref{fig:background}(c) shows that if all loads were to have zero verification latency, we would get a $5.7\times$ speedup.
    
    The simplest solution would be replacing all 2D YSCs with 1D YSCs, as 1D YSCs
    effectively have zero additional verification latency.
    However, 1D YSCs are less storage efficient than 2D YSCs, so this solution is undesirable.
    The goal of this work is to design an FTQC memory hierarchy that achieves the low verification latency of 1D YSCs
     while retaining the storage benefits of 2D YSCs.
     We discuss our methodology before presenting our solution.
     
\ignore{
    \begin{figure}[h]
        \centering
        \includegraphics[width=1\columnwidth]{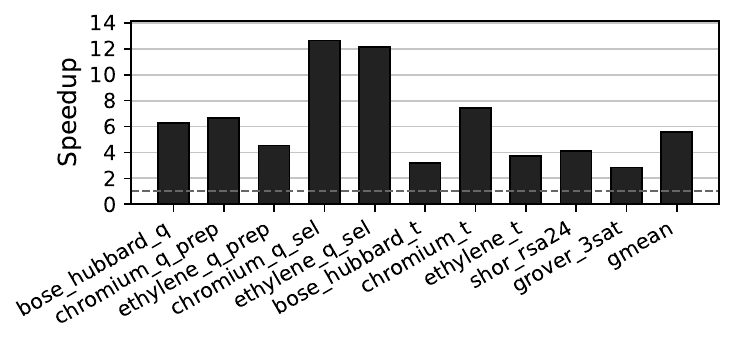}
        \caption{Speedups for an ideal memory with no verification latency. The gmean speedup is $5.6\times$.}
        \label{fig:ideal}
    \end{figure}
}
    
\ignore{
    A load involves executing lattice surgery measurements to measure
    the $ZZ$ operator between a program qubit within the YSC memory and a surface code
    logical qubit prepared in the $\ket{+}$ state.
    However, as the program qubit is encoded using a YSC, the $Z$ logical operator of
    the program qubit spans four different surface code qubits resident within the YSC.
    Figure~\ref{fig:2d_ysc_load} shows how a program qubit is loaded from a YSC memory
    into a surface code logical qubit (``\textit{LD}").
    In Figure~\ref{fig:2d_ysc_load}, the four qubits comprising the program qubit's
    logical $Z$ operator are denoted ``Z".
    In steps one through three, \textit{LD} is entangled with each 
    of the $Z$ operator qubits.
    Once the $ZZ$ measurement completes, \textit{LD} and the program qubit become
    entangled, but the original memory location within the YSC must be destroyed 
    to complete the load.
    To do so, we must measure the original program qubit in the $X$ basis.
    Like with the $Z$ operator, there are four surface code qubits within the YSC
    that correspond to the program qubit's $X$ operator.
    These surface code qubits are denoted ``X" in Figure~\ref{fig:2d_ysc_load}.
    As shown in steps four through six,
    the four $X$ operator qubits perform $XX$ measurements with an ancilla 
    surface code logical qubit prepared in the $\ket{0}$ state.
    Then, this ancilla qubit is measured in the $X$ basis, 
    which destroys the original memory location.
    Now, \textit{LD} contains the desired program qubit.
    In total, a load operation lasts eight cycles.

    \begin{figure}[!htb]
        \centering
        \includegraphics[width=\columnwidth]{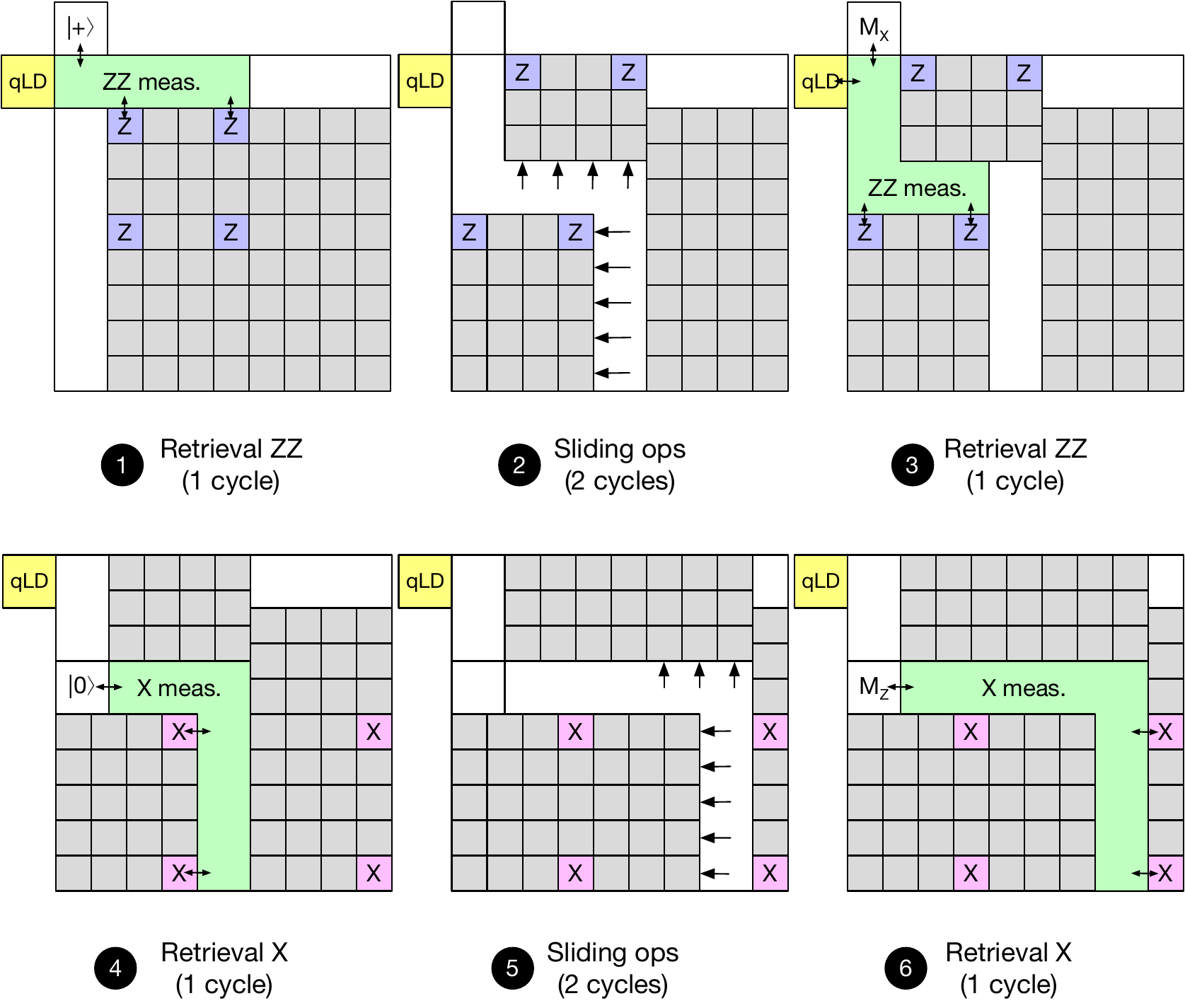}
        \caption{The series of lattice surgery operations required to load
                a program qubit from a 2D YSC block.
                The overall latency of a load is eight cycles.}
        \label{fig:2d_ysc_load}
    \end{figure}
}

\ignore{
\subsection{Load Verification}
    Unfortunately, loading to, or storing from, a YSC is not fault-tolerant.
    During memory access, an error can occur on any of the $X$ or $Z$ operator
    qubits involved in a memory access.
    As the YSC's yokes cannot be measured during a memory access,
    these surface code qubits are unprotected by the yokes.
    Thus, if any error cannot be corrected by the constituent surface codes 
    themselves, a logical error will occur, resulting in the 
    loaded/stored program qubit
    having an erroneous state~\cite{gidney_stack_exchange_yoke_blocking}.
    Recovering from this error requires measuring the yokes of the YSC after
    the memory access to accurately determine changes to the 
    program qubit's logical state.
    We refer to this process as \textit{verification}.
    The verification latency depends on how long it takes for the YSC to measure
    all of its yokes.
    For a 2D YSC, the verification latency is 178 cycles.

    Until the YSC completes the verification procedure, the loaded/stored program
    qubit's syndrome history is unresolved.
    If a program qubit's syndrome history is unresolved, then it cannot
    execute any non-Clifford operations.
    Thus, store verification does not stall program execution as no compute
    instructions are dependent on the outcome of a store.
    In contrast, load verification does stall program execution.
    While the FTQC can attempt to amortize the verification latency
    by batching multiple loads together,
    our studies suggest that even with such optimizations, load verification
    causes a $5.6\times$ slowdown relative to an idealized system with
    no verification latency.
}

\ignore{
\subsection{Goal}
    Load verification latency degrades the performance of the
    FTQC's memory subsystem ($5.6\times$ slowdown).
    The goal of this paper to mitigate the impact of verification on program performance.
    To this end, we propose {\textit{PROMPTLY}}.
}

\ignore{
\subsection{Basics of Quantum Error Correction}
    A QEC code constructs reliable logical qubits by encoding them across many noisy physical qubits. Specifically, an
    $[[n,k,d]]$ code encodes $k$ logical qubits into $n$ physical qubits with code distance $d$, with a \textit{code
    rate} of $k/n$. To monitor the encoded state, the code repeatedly measures parity checks called
    \textit{stabilizers}, whose outcomes indicate whether the state has suffered an error. The collection of outcomes
    from measuring all stabilizers forms a \textit{syndrome}, and the sequence of operations used to obtain this
    syndrome comprises a \textit{syndrome extraction round}. A QEC \textit{decoder} processes these syndromes, infers
    the most likely error and determines the corresponding correction.
    
    QEC codes can be \textit{concatenated} such that the logical qubits of one code (the \textit{inner code}) serve as
    the data qubits of a second code (the \textit{outer code}). The inner code suppresses physical faults within each
    encoded block, while the outer code provides protection against residual logical faults of the inner code. By
    protecting information at two levels, the concatenated construction achieves higher reliability than either code
    alone. This improvement can be described in terms of an effective code distance $d_{\text{eff}}$, which typically
    scales with the product of the constituent code distances, $d_{\text{eff}}=\mathcal{O}(d_{\text{in}}\cdot
    d_{\text{out}})$ \cite{poulin2006optimal}.
    
    An $[[n,k,d]]$ quantum error correcting (QEC) code encodes $n$ noisy data qubits into $k$ reliable logical qubits.
    The code distance $d$ parameterizes the reliability of this code block, and a distance $d$ block can correct up-to
    $(d-1)/2$ errors on its constituent data qubits.
    An $[[n,k,d]]$ code has $n-k$ checks that are periodically measured to detect errors on the data qubits.
    The results of these check measurements form a bitstring known as a \textit{syndrome}.
    Then, this syndrome is processed by a classical \textit{decoder}, which identifies any changes to the logical state.
    These changes are then committed into the corresponding logical qubit's \textit{Pauli frames}, which are two bits
    that
    track $X$ and $Z$ errors on the logical qubit, respectively \cite{knill2007quantum}.

    QEC codes can be \textit{concatenated} such that the logical qubits of one code (the \textit{inner code}) serve as
    the data qubits of a second code (the \textit{outer code}). The inner code suppresses physical faults within each
    encoded block, while the outer code provides protection against residual logical faults of the inner code. By
    protecting information at two levels, the concatenated construction achieves higher reliability than either code
    alone. This improvement can be described in terms of an effective code distance $d_{\text{eff}}$, which typically
    scales with the product of the constituent code distances, $d_{\text{eff}}=\mathcal{O}(d_{\text{in}}\cdot
    d_{\text{out}})$ \cite{poulin2006optimal}.
}

\ignore{
    In this paper, we primarily study the \textit{surface code} \cite{fowler2012surface, litinski2019game}.
    The surface code is an $[[d^2, 1, d]]$ QEC code with a grid structure, such that the $X$ and $Z$ checks
    of the code are incident at most four qubits.
    Furthermore, the data qubits on the sides of the surface code correspond to logical operators.
    In this paper, we use the convention where the data qubits on the top and bottom sides of the
    surface code implement the logical $Z$ operator, and the data qubits on the left and right sides
    implement the logical $X$ operator.
    Typically, physical implementations of the surface code use $2(d+1)^2$ physical qubits.
    $d^2$ of these physical qubits are the data qubits of the code, and
    the remaining qubits are used for check measurements and for faciltiating logical operations.
    Figure {\color{red} XX} presents an example of a $d = 5$ surface code as well as its
    $X$ and $Z$ logical operators.

    Logical operations on the surface code are performed using \textit{lattice surgery}.
    Lattice surgery facilitates Pauli-product measurements between two or more surface
    code logical qubits (i.e., $XX$ or $ZZ$) by connecting the logical operators of the
    involved logical qubits qubits using ancilla data qubits, $X$, and $Z$ checks.
    Figure {\color{red} XX} shows an example of surface code lattice surgery between
    two logical qubits.
    After $d$ or more rounds of measuring these intermediate checks, the intermediate data qubits
    are measured. 
    The product of these intermediate data qubit measurements and some of the intermediate
    checks form the lattice surgery measurement outcome. Accordingly, logical operations are naturally scheduled at a
    granularity of $d$ rounds. Thus, $d$ syndrome extraction rounds constitute one \textit{logical cycle} of the surface
    code.
    After the measurements, all involved logical qubits are in an entangled state dictated by the 
    lattice surgery measurement outcome. 
    
    As physical readout is unreliable, resolving a lattice surgery measurement accurately
    requires using the decoder to accurately resolve the lattice surgery measurement outcome.
    Furthermore, as the measurement result depends on the Pauli frames of the involved
    logical qubits, the decoder must also resolve the syndrome history prior to the lattice
    surgery operation before resolving the measurement outcome.
}

\ignore{
    The main drawback of the surface code is its high spatial cost, driven primarily by its low code rate, which scales as
    $1/d^2$. To store logical information more densely, the yoked surface code concatenates the surface code with a
    high-rate outer code \cite{gidney2025yoked}. In this construction, the outer code imposes additional parity checks
    called \textit{yokes} across the logical operators of the inner surface code patches. These yokes are measured using
    lattice surgery. We refer to one full measurement of all yokes in a block as a \textit{yoke-check sequence}.

    Decoding in the yoked surface code combines violated yokes with syndrome information from the inner surface code
    patches. This \textit{soft information} captures the quality of each patch, i.e., the likelihood that it has suffered a
    logical error. A violated yoke check indicates that a patch-level logical error has occurred, while the patch-quality
    information is used to localize and correct the error. Thus, a logical error in the yoked surface code must span the
    logical support of multiple surface-code patches, increasing the effective code distance without increasing the distance
    of individual patches. Yoked surface codes therefore improve the density of stored logical information without requiring
    additional hardware connectivity beyond planar nearest-neighbor interactions.

    Depending on the choice of the outer code, there are two primary families of yoked surface codes:

    \begin{enumerate}
        \item \textbf{1D yoked surface code}: Uses a $[[n,n-2,2]]$ code, where $n$ is even \cite{steane1996simple}. The $n$
        surface code patches are arranged linearly. The outer code has two weight-$n$ yokes: one acting on logical $Z$
        operators of all patches (as shown in Figure \ref{fig:1d_yoked_code}) and the other on their logical $X$ operators.
        
        \item \textbf{2D yoked surface code}: Uses a $[[n^2, n^2-4n+2, 4]]$ code, where $n$ is divisible by 4. The $n \times
        n$ surface code patches are arranged in a square grid. The outer code has $4n-2$ weight-$n$ yokes. To minimize
        spatial overhead, the ancilla space used for lattice surgery is shared across yoke checks, so these checks must be
        performed sequentially.
    \end{enumerate}

    \begin{figure}[h]
        \centering
        \includegraphics[width=0.5\linewidth]{figures/1D Yoke Check.png}
        \caption{$Z^{\otimes10}$ parity check on a $[[10,8,2]]$ yoked surface code}
        \label{fig:1d_yoked_code}
    \end{figure}
}
\ignore{
\subsection{Loading from Yoked Surface Codes}
    The 2D yoked surface code improves logical storage density by roughly \(3\times\), making it attractive as a
    high-density memory \cite{gidney2025factor, khattar2025verifiable}. We now describe how a logical qubit encoded in a 2D
    block can be loaded out.

    In yoked surface codes, the $X$ and $Z$ logical operators of the encoded qubits are distributed across multiple
    constituent surface code patches---across two patches in the 1D code and four patches in the 2D code
    \cite{gidney2025yoked}. We can perform lattice surgery with a logical operator of the yoked surface code by
    simultaneously stitching to the $X$ or $Z$ operators of the constituent surface code patches. To match the error rate of
    memory, such operations must be performed at the effective distance of the yoked surface codes. Accordingly, the surgery
    hallway must have width and height \(d_{\mathrm{eff}}\), and its intermediate checks must be measured for
    \(d_{\mathrm{eff}}\) syndrome extraction rounds. Therefore, one logical cycle of a yoked region consists of
    \(d_{\mathrm{eff}}\) rounds, matching the cycle length of an unyoked surface code at the same distance.

    As an illustrative example, Figure \ref{fig:yoked_memory_load} shows the steps required to load a logical qubit $q$ from
    a $[[64,34,4]]$ yoked surface code block into an external surface code patch. The load uses the distributed logical
    operators of \(q\):

\begin{itemize}
    \item Steps 1,3 perform two lattice surgery operations to measure the joint parity between the logical \(Z\) operator of the external
    patch and the logical \(Z\) operator of \(q\), which is itself the product of \(Z\) operators on four constituent surface code patches.
    \item Steps 2,5 perform sliding operations to reposition the ancilla hallways adjacent to the patches supporting the logical operators of \(q\) \cite{dubey2025access}.
    \item Steps 4,6 perform two lattice surgery operations to measure \(q\) in the logical \(X\)-basis by stitching to the \(X\) operators of
    its constituent patches.
\end{itemize}

\begin{figure*}
    \centering
    \includegraphics[width=1\linewidth]{figures/Yoked Memory Load.png}
    \caption{Loading from a 2D Block}
    \label{fig:yoked_memory_load}
\end{figure*}

    Following this sequence, the state of $q$ is loaded to the external surface code patch. As shown, a load requires 8
    cycles. Conversely, storing a qubit from an external patch into the 2D yoked block requires two lattice surgery
    operations together with sliding, and takes 4 cycles.
}\ignore{
\subsection{The 2D Memory Bottleneck}

In architectures that use an unyoked compute region with 2D yoked memory, qubits are loaded out of 2D memory when they
are needed for computation. This can be done using the load sequence described above. However, completing the load does
not make the qubit immediately ready for arbitrary use. An additional latency is encountered before that qubit regains
the full effective protection of the yoked surface code.

This delay arises because errors on the constituent surface code patches of the 2D block are detected and corrected only
when the yoke checks are performed. Between successive yoke check sequences, each inner patch is exposed at roughly the
logical error rate as an unyoked patch of the same inner distance. If one of the patches supporting the logical operator
used in the load sequence suffers a logical error, that fault corrupts the joint parity measured during lattice surgery
and is propagated to the external surface code patch. Importantly, this remains a detectable event for the yoked code:
since logical operators commute with the yoked stabilizers, the erroneous state violates an outer code stabilizer that
will be identified when the next yoke check sequence completes. At that point, the faulty patch, and hence the loaded
qubit, can be corrected.

Consequently, a qubit loaded from 2D memory is available in compute immediately, but it does not initially have the full
fidelity of the yoked code. Until the next yoke check sequence completes, its reliability is limited by the supporting
inner patches. The inner decoder may also report low-confidence evidence on one or more of these patches; we refer to
such evidence as a \textit{suspicious event}. Figure \ref{fig:sus_event} shows that these suspicious events are highly
likely; if the desired error-rate per logical cycle is $10^{-15}$, then a $d=11$ surface code patch has a suspicious
event in $68\%$ of cycles. Because any suspicious event on a supporting patch lowers confidence in the loaded state, a
conservative architecture treats a qubit loaded from 2D storage as \textit{not ready} until the next yoke check sequence
resolves these uncertainties.

\begin{figure}[h]
    \centering
    \includegraphics[width=1\linewidth]{figures/gap_threshold_tradeoff.png}
    \caption{Probability of suspicious events per-cycle vs logical error rate threshold}
    \label{fig:sus_event}
\end{figure}

This delayed readiness is acceptable for Clifford computation. If a later yoke check reveals that the loaded qubit
carried a Pauli error, the effect can be propagated retrospectively by updating the software-tracked Pauli frame of the
qubits that were operated on. But if a non-Clifford operation uses an unresolved loaded state, and a later yoke check
reveals that the state was incorrect, then the computation would have to be undone and recomputed from that operation.
Such rollback is unsafe in general, and may be impossible after irreversible operations such as logical measurements.
The safe architectural choice is therefore to stall until the next yoke-check sequence completes before allowing the
loaded qubit to participate in a non-Clifford operation.

This waiting time is central to the bottleneck of 2D yoked memory. A 2D yoked block contains many outer-code
stabilizers, and these checks must be performed sequentially. As a result, one full yoke check sequence can span
hundreds of logical cycles. Every load from 2D memory that needs a non-Clifford operation may therefore incur a long
readiness delay, causing 2D storage to behave as a high-latency memory and throttling overall execution.
}

\section{Methodology}

\subsection{Compilation}

Our compiler takes benchmark programs provided as QASM files and translates them into Clifford+$RZ$ operations. 
We synthesize $RZ$ gates into a sequence of Clifford+$T$ gates using Gridsynth~\cite{ross2014_gridsynth}.
These syntheses are further optimized using gate coalescing~\cite{wang2024_optimizing_ftqc}.
Finally, our compiler inserts load/store instructions to move program qubits between the compute
and memory subsystems of the FTQC in a manner similar to prior works such as HetEC, LSQCA, 
and qSIEVE~\cite{viszlai2026qsieve, kobori2025_lsqca, stein2025_hetec}.

\subsection{Simulation}
    
    We use the compiled binaries generated by our compiler to drive a cycle-level simulator that
    simulates program execution on an FTQC.
    Our simulator models the compute subsystem, memory subsystem, and magic state production
    in detail, including routing space contention in the compute and memory subsystems. In particular, we allow at most one memory operation to execute at any time, so all loads and stores are serialized.
    Table~\ref{tab:three_level_config} shows the simulation configuration used in
    our evaluations which models a three-level memory hierarchy whose intermediate memory tier
    contains a \textit{single} 1D YSC with capacity eight.
    Note that the baseline system, whose memory subsystem only contains 2D YSCs, does not
    contain this intermediate memory tier.
    This simulation configuration reflects an FTQC operating in the teraquop regime, 
    which corresponds to a logical error rate per syndrome extraction round
    of $10^{-15}$.
         
\ignore{
The memory subsystem is composed of YSC blocks. Each block must measure its yokes periodically, since the interval between these checks controls the logical error rate of the block. 

A 2D YSC block alternates between two phases: servicing memory operations and measuring all of its yokes. During the servicing phase, our system batches pending loads and stores to the block to amortize the verification latency of multiple successive memory accesses. If a 2D block receives no memory requests for 14 cycles, it transitions to a yoke check phase, during which its yokes are measured serially without interruption. To bound the time spent in the servicing phase, at most 8 memory operations may be served before the yokes are checked, thereby bounding the resulting error rate. 

In contrast, a 1D YSC block can interleave its yoke checks with memory operations. As a result, an explicit check of the yokes is triggered only when the block remains idle for 100 cycles.
}

    \begin{table}[!htb]
        \begin{center}
            \footnotesize
            \caption{Simulation Configuration}
            \label{tab:three_level_config}
            \begin{tabular}{|c|c|}
                \hline
                \multicolumn{2}{|c|}{\textbf{Compute Subsystem}} \\
                \hline
                \hline
                Capacity (Logical Qubits) & 4 \\
                Distance & $d=25$ \\
                Physical Qubits per Logical Qubit & 2028 (incl. Routing) \\
                Logical Error Rate & $d \times 10^{-15}$ \\ 
                \hline
                \hline
                \multicolumn{2}{|c|}{\textbf{L1 Memory (1D Yoked Surface Code) -- \textit{not in baseline}}} \\
                \hline
                \hline
                Surface Code Count & 10\\ 
                Capacity (Logical Qubits) & 8 \\
                Inner Surface Code Distance & $d=15$ \\
                Retrieval Latency & 8 cycles \\
                Verification Latency & 0 cycles \\
                Physical Qubits per Logical Qubit & 1100 (incl. Routing) \\
                \hline
                \hline
                \multicolumn{2}{|c|}{\textbf{L2 Memory (2D Yoked Surface Codes)}} \\
                \hline
                \hline
                Surface Code Count & 256 \\
                Capacity (Logical Qubits) & 194  \\
                Inner Surface Code Distance & $d=11$ \\
                Retrieval Latency & 8 cycles \\ 
                Verification Latency & 178 cycles \\
                Physical Qubits per Logical Qubit & 454 (incl. Routing)  \\
                \hline
                \hline
                \multicolumn{2}{|c|}{\textbf{Magic State Production}} \\
                \hline
                \hline
                Physical Qubit Footprint & 47.8K \\
                $d=3$ Cultivation Units \cite{gidney2024_cultivation} & 62 \\
                $15:1$ Distillation Units \cite{litinski2019_msd} & 6 \\
                Magic State Infidelity & $10^{-12}$ \\
                \hline  
            \end{tabular}
        \end{center}
    \end{table}

\subsection{Evaluated Workloads}

We perform our evaluations using ten diverse, large-scale benchmarks that represent promising applications for FTQCs. Table \ref{tab:workloads} summarizes the characteristics of these benchmarks, namely their qubit counts, instruction counts, percentage of $T$ gates (T\%), and percentage of memory accesses (M\%) in the compiled binary, and speedup of an idealized system with zero-cycle verification.
Our workload suite contains three types of applications: 
block-encoding (Qubitization)~\cite{low2019_qubitization,
                                    gilyen2019_qsvt}, 
Trotterization~\cite{suzuki1976_trotter_suzuki_decomp},
and algorithmic benchmarks such as Shor's~\cite{shor1994_algorithms} 
or Grover's~\cite{grover1996_algorithm}. 
We use Hamiltonians from HamLib~\cite{sawaya2024_hamlib} to construct 
each block-encoding and Trotterization benchmark.
For each Hamiltonian, we denote the block-encoding benchmarks with
a ``q" suffix and the Trotterization benchmarks with a ``t"
suffix.
For each benchmark, we simulate the first 100M instructions and 
measure performance using \textit{Instructions per Cycle (IPC)}.

Within a block encoding benchmark, there are two prominent subroutines:
\texttt{PREPARE} and \texttt{SELECT}.
Both subroutines exhibit very different program behavior.
While \texttt{PREPARE} is $RZ$-heavy and thus compute intensive,
\texttt{SELECT} is highly memory intensive.
Thus, we evaluate the \texttt{PREPARE} and \texttt{SELECT} subroutines
separately, and they are denoted by the ``prep" and ``sel" suffixes,
respectively.
The only exception to this rule is for the Bose-Hubbard benchmark.
The benchmark is so short that both \texttt{PREPARE} and 
\texttt{SELECT} appear in the first 100M instructions of the benchmark.

\begin{table}[!htb]
    \begin{center}
        \footnotesize
        \caption{Evaluated Benchmarks}
        \label{tab:workloads}
        \begin{tabular}{|c||c|c|c|c|c|}
            \hline
            Workload & Qubits & Inst. & T\% & M\% & Ideal Speedup \\
            \hline
            \hline

            bose\_hubbard\_q        & 2228 & 204M & 75.0\% & 6.7\% & $5.96\times$ \\
            chromium\_q\_prep        & 167 & 471M & 76.4\% & 4.7\% & $6.58\times$ \\
            ethylene\_q\_prep        & 289 & 1.3B & 79.4\% & 3.1\% & $5.67\times$ \\
            chromium\_q\_sel        & 167 & 292M & 20.1\% & 33.0\% & $12.5\times$ \\
            ethylene\_q\_sel        & 289 & 436M & 21.4\% & 24.4\% & $12.4\times$ \\
            \hline
        
            bose\_hubbard\_t        & 2188 & 161M & 93.1\% & 1.3\% & $3.21\times$ \\
            
            chromium\_t         & 121 & 340M & 75.2\% & 7.5\% & $7.42\times$ \\
            ethylene\_t         & 241 & 320M & 88.7\% & 2.3\% & $3.89\times$ \\
            \hline

            shor\_rsa24    & 111 & 174M & 86.6\% & 1.7\% & $4.12\times$ \\
            grover\_3sat    & 1710 & 295M & 91.2\% & 1.7\% & $2.83\times$ \\
            \hline
        \end{tabular}
    \end{center}
\end{table}

\subsection{Application Fidelity Estimation}

    We compute application fidelity by estimating the fidelity
    of the compute and memory subsystems separately.
    As we evaluate a subset of a program (100M instructions),
    we scale statistics output by our simulator, such as the
    number of cycles, to the total number of instructions
    in the program.

    \subsubsection{Compute Subsystem Fidelity}
    The fidelity of the compute subsystem relies on the following:
    (1)~the number of compute subsystem cycles ($C$),
    (2)~the capacity of the compute subsystem ($N$),
    (3)~the per-cycle error rate of the surface code ($e_C$),
    (4)~the number of $T$ gates in the program ($T$),
    and (5) $T$ infidelity ($e_T$).
    $N$, $e_C$, and $e_T$ are shown 
    in Table~\ref{tab:three_level_config}, whereas $C$ and $T$
    are calculated after simulation.
    Equation~\eqref{eqn:cs_fidelity} shows how these variables
    determine the fidelity of the compute subsystem \cite{huggins2025fluid}.
    \begin{equation}
    \label{eqn:cs_fidelity}
        F_{\mathrm{compute}} = (1-e_C)^{N \times C} \cdot (1 - e_T)^T
    \end{equation}
    
    \subsubsection{Memory Subsystem Fidelity}
    
    The fidelity of the memory subsystem is determined by the error rates of its YSCs. The measurement of all yokes in a YSC completes one \emph{yoke cycle}. The logical error probability of a yoke cycle  depends on the following: (1) the distance of the constituent surface code patches ($d_{\mathrm{in}}$), (2) the number of surface code patches ($n$), and (3) the number of syndrome extraction rounds that elapse during the yoke cycle ($r$). The logical error probabilities of a 1D YSC and a 2D YSC yoke cycle are given by Equation~(\ref{eq:ysc-error}) \cite{gidney2023_yoked_sc}. 
        \begin{equation}
        \label{eq:ysc-error}
        \begin{split}
        e_{\mathrm{1D}} &= r^2\cdot n\cdot 10^{-d_{\mathrm{in}}}/2000 \\
        e_{\mathrm{2D}} &= r^4\cdot n^2 \cdot 200^{-d_{\mathrm{in}}}/20000 \\
        \end{split}
        \end{equation}

    The dependence on $r$ reflects the need to measure yokes periodically to prevent errors from accumulating on the constituent surface codes of a YSC. During simulation, we allow $r$ to vary between different yoke cycles according to the memory access pattern of the program. To bound $r$, we allow a 2D YSC to serve memory accesses for at most 100 consecutive cycles. It also begins measuring its yokes after 14 consecutive cycles without a memory access. In contrast, a 1D YSC can interleave yoke measurements with memory accesses. We therefore check its yokes on every memory access or after 50 consecutive idle cycles. Consequently, a high memory-access rate delays yoke measurements and increases the logical error probability for a 2D YSC, but not that of a 1D YSC. 
    
    Since the duration of a yoke cycle may vary during execution, so may its logical error probability. Let $e_y$ denote the error probability of the $y^{\mathrm{th}}$ yoke cycle, calculated using its realized value of $r$. Then, the total fidelity of a YSC that completes $Y$ yoke cycles during execution is calculated using 
    Equation (\ref{eq:ysc-accumulate}). Equation (\ref{eq:ms_fidelity}) accumulates the fidelity across all $B$ YSC blocks in the memory subsystem.

    \begin{align}
        \label{eq:ysc-accumulate}
        F_{\mathrm{YSC}} &= \prod_{y=1}^Y(1-e_y) \\
        \label{eq:ms_fidelity}
        F_{\mathrm{memory}} &= \prod_{b=1}^B F_{\mathrm{YSC, b}}
    \end{align}

    \subsubsection{Overall Fidelity}

    Assuming independent errors, the overall application fidelity is the product of compute system fidelity and memory system fidelity.

    \begin{equation}
        F_{\mathrm{overall}} = F_{\mathrm{compute}}\cdot F_{\mathrm{memory}}
    \end{equation}
    
\section{Trident: Towards a Three-Level Hierarchy}

To achieve low verification latency overheads while maintaining the high density of 2D YSC, we propose {\em Trident}. Trident is based on the observation that the effective verification latency for 1D YSC is zero as the verification procedure can be interleaved with the retrieval of the logical qubit. However, using 1D YSCs as a memory in place of 2D YSCs is undesirable as 2D YSCs are much more storage efficient than 1D YSCs ($4.5\times$ better than surface code vs $1.8\times$). Hence, we must devote most of the physical qubits to the 2D YSC memory for getting high density.

The main insight in Trident is to augment the two-level memory hierarchy with an intermediate memory tier consisting of 1D YSC memory. If a majority of loads are from the 1D YSC memory (L1) rather than the 2D YSC memory (L2), then the average load verification latency will be much less than 178 cycles. Trident expands the traditional two-level memory hierarchy into a three-level memory hierarchy. Trident uses a single, 8-entry 1D YSC block in the L1 memory, which consumes 8.8k physical qubits total, including routing space.  The effectiveness of Trident depends on the hit rate of the L1 tier, so it is important to manage the L1 space intelligently.  We propose two variants of Trident. The first, {\em Trident-D}, manages the L1 as a cache in a demand-driven manner. The second, {\em Trident-P}, manages the L1 as a prefetch buffer in a prefetch-driven manner. In this section, we first discuss the load operation in Trident and then discuss the design details and effectiveness of Trident-D.

\subsection{Trident: Design and Operation}
    \label{sec:load_semantics}

\begin{figure}[!htb]
        \centering
        \includegraphics[width=\columnwidth]{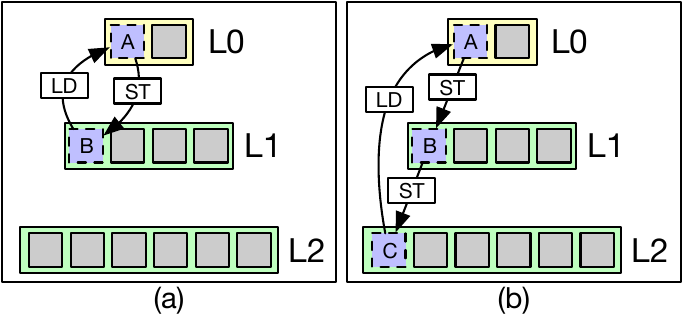}
        \caption{Design and Operation of Trident.  (\textbf{a})~Semantics of loading qubit \texttt{B} from the L1 and into 
                    the L0 (compute subsystem).
                    Qubit \texttt{A} must be stored to make space for \texttt{B}.
                  (\textbf{b})~Semantics of loading qubit \texttt{C} from the L2 and into the L0.
                  To make space for \texttt{C}, \texttt{A} must be stored into the L1.
                  Subsequently, to make space for \texttt{A}, \texttt{B} must be stored
                  into the L2 location made vacant by \texttt{C}.
                  }
        \label{fig:load_semantics}
    \end{figure}

    Figure~\ref{fig:load_semantics} shows an overview of Trident.  Due to the no-cloning theorem, a quantum memory hierarchy must be exclusive:
    each program qubit can only occupy a single memory location in the memory
    hierarchy.
    Consequently, there is an aspect of ``conservation" when performing
    memory accesses.
    A load, for example, clears a location in the L1 or L2 memories and occupies
    a location in the L0 (compute subsystem).
    However, as the L0 will be full, space must be allocated for the loaded
    program qubit.
    Thus, each load, regardless of the source memory location, must
    have a corresponding store.
    
    Trident provides two load semantics for moving data throughout the
    three-level memory hierarchy.
    Figure~\ref{fig:load_semantics}(a) shows how Trident handles a load
    to the L1 memory.
    As shown, when qubit \texttt{B} is loaded into the L0, Trident removes
    some qubit \texttt{A} from the L0 to allocate space for \texttt{B}.
    Subsequently, Trident stores \texttt{A} into the memory location vacated by \texttt{B}
    in the L1.
    Figure~\ref{fig:load_semantics}(b) shows how Trident manages a load
    from the L2 memory.
    When qubit \texttt{C} is loaded into the L0, Trident removes a qubit \texttt{A}
    from the L0. If Trident elects to store \texttt{A} into the L1, Trident must evict some qubit \texttt{B} 
    residing within the L1.
    Then, Trident stores \texttt{A} into the location left vacant by \texttt{B} in the
    L1 and stores \texttt{B} into the location left vacant by \texttt{C} in the L2.  Alternatively, Trident may store the removed qubit from L0 directly into the L2.

    There are more options for data movement than exposed by Trident. However, in our studies, additional load semantics beyond the two described
    above provided negligible benefit.
    So, we did not include them.

    \begin{figure*}[!htb]
        \centering
        \includegraphics[width=\linewidth]{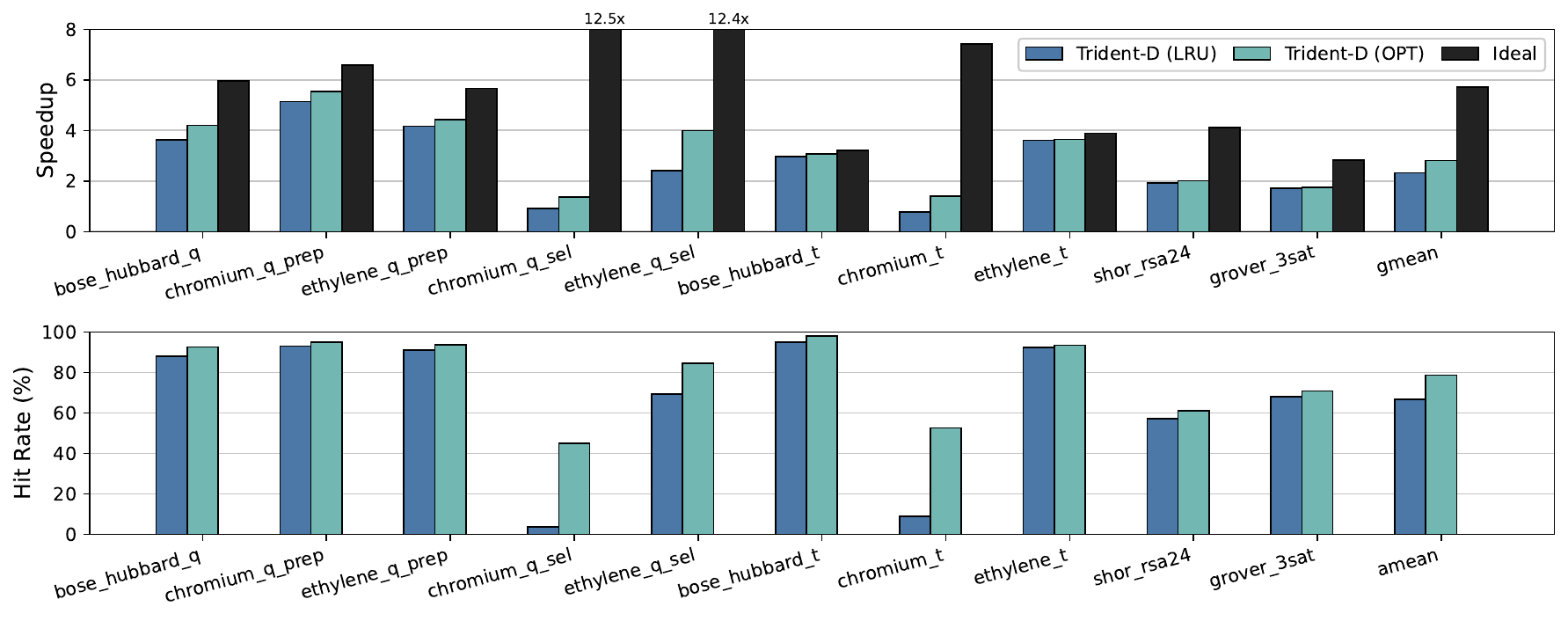}
        \caption{\textbf{(Top)} Speedups for Trident-D with LRU, Trident-D with OPT, and an idealized FTQC with
                zero verification latency.
                Trident-D with LRU, Trident-D with OPT, and the idealized
                FTQC achieve speedups of $2.3\times$, $2.8\times$, and $5.7\times$, respectively.
                \textbf{(Bottom)} L1 hit rates for Trident-D with LRU and Trident-D with OPT, which
                    are, on average, 66.7\% and 78.7\%, respectively.}
        \label{fig:lru_opt_speedup}
    \end{figure*} 
  
\subsection{Trident-D: Design}
\begin{figure}[!htb]
        \centering
        \includegraphics[width=\linewidth]{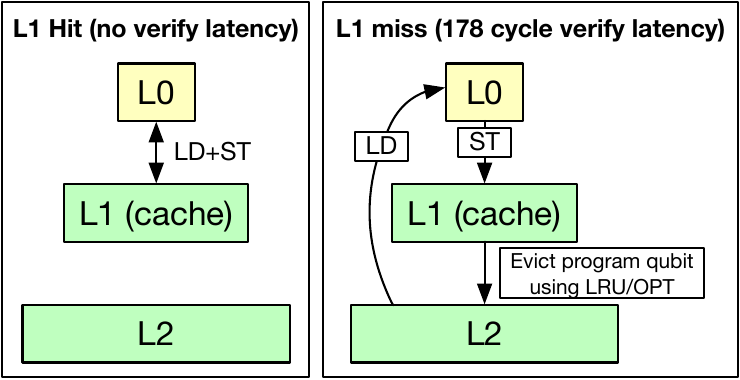}
        \caption{Trident-D operates the L1 as a cache.
                On an L1 miss, Trident-D evicts one of the program qubits
                resident within the L1 using either LRU or OPT (Compiler-managed).}
        \label{fig:trident_d_overview}
    \end{figure}

    Trident provides a framework for operating a three-level memory
    hierarchy on surface code FTQCs.
    However, Trident does not specify the specific decisions
    that should be taken during a load, such as which qubit to
    store to a lower memory tier.
    To this end, we propose \textit{Trident-D (Demand-Driven Trident)}
    which operates the three-level memory hierarchy as a demand-driven
    hierarchy.
    A high-level overview of Trident-D is shown in Figure~\ref{fig:trident_d_overview}.

    Program qubits that are accessed more frequently are
    preferentially stored in the L1 memory, whereas infrequently accessed
    program qubits are stored in the L2 memory.
    In other words, Trident-D operates the L1 as a cache for the L2.

    Trident-D is a \textit{compile-time} solution that modifies a program
    already compiled to a two-level memory hierarchy, which we obtain by
    using the algorithms proposed in prior works~\cite{stein2025_hetec, kobori2025_lsqca, viszlai2026qsieve}.
    Trident-D adds additional instructions to the input program to manage
    the L1 memory.
    Trident-D can use either LRU or Belady's optimal replacement
    policy (OPT).
    We discuss both implementations of Trident-D.

    \subsubsection{Trident-D with LRU}
        Trident-D reads the program and tracks the contents of
        the L1 memory.
        When a program qubit is stored from the L0 and into the
        L1, Trident-D records the time at which this qubit
        was stored.
        This timestamp will be used later for eviction decisions.
        Each load in the initial quantum program will either
        target a program qubit currently within the L1 or within
        the L2.
        If the load accesses the L1 (cache hit), then Trident-D
        does not change the program as the load and its
        corresponding store obey the first load semantic.
        However, if the load accesses the L2 (cache miss), then
        Trident-D must identify which qubit in the L1 to
        evict and store to the L2 as per the second load semantic.
        Trident-D with LRU evicts the oldest qubit
        in the L1\footnote{Note that as the memory hierarchy is entirely
                            exclusive, every hit removes the accessed qubit
                            from its original memory location, so LRU replacement degenerates into 
                            FIFO replacement.}.

    \subsubsection{Trident-D with OPT}
        While OPT replacement is not practical for hardware-managed caches, 
        in our case, we have a compiler managed cache, which has knowledge of future accesses to each qubit.  
        Trident-D with OPT functions similarly to Trident-D with LRU.
        The only difference is that to implement OPT, we must record when each qubit is accessed.  
        To accomplish this, we read the input program and record
        the DAG layer of each load as a timestamp.
        The program is not modified in this first read.
        Then, Trident-D re-reads the entire program with the intent
        to modify it.
        For an L2 load, Trident-D with OPT uses the pre-recorded timestamps
        from its first read of the program to identify the L1 program
        qubit whose next load is farthest into the future.
        The evicted qubit is then stored into the L2 as per
        the second load semantic.

\ignore{
    As an input, Trident-D operates on a program already compiled
    to a two-level memory hierarchy, which we obtain by
    using algorithms proposed in prior 
    works~\cite{stein2025_hetec, viszlai2026qsieve, kobori2025_lsqca}.
    Trident-D reads the program DAG of the input program and
    records the DAG layer of each load as a timestamp.
    The program is not modified in this first read.
    
    Trident-D re-reads the program again with the intent
    of modifying the program.
    During this re-read, Trident-D tracks the contents
    of the L1 memory.
    Once Trident-D finds a load instruction, it first
    checks if the load accesses a program qubit currently
    present in the L1.
    If so, Trident-D does not need to modify the program
    as the load and its corresponding store obey the
    first load semantic.
    However, if the load accesses a program qubit in the
    L2, Trident-D must identify which qubit in the L1
    to evict and store to the L2 as per the second load semantic.
    Trident-D evicts the qubit whose load is farthest in
    the future.
    Additionally, Trident-D modifies the program to
    implement this eviction.
}

\subsection{Trident-D: Results}
    Figure~\ref{fig:lru_opt_speedup} (top) shows the speedups of Trident-D
    with LRU, Trident-D with OPT, and an idealized two-level hierarchy
    with zero verification latency.
    As shown, Trident-D with LRU, Trident-D with OPT, and the idealized
    FTQC achieve speedups of $2.3\times$, $2.8\times$, and $5.7\times$,
    respectively.
    The speedup of Trident-D with OPT is slightly larger than Trident-D
    with LRU, which is expected as LRU is an approximation of OPT.

    \begin{figure*}[!htb]
        \centering
        \includegraphics[width=0.95\textwidth]{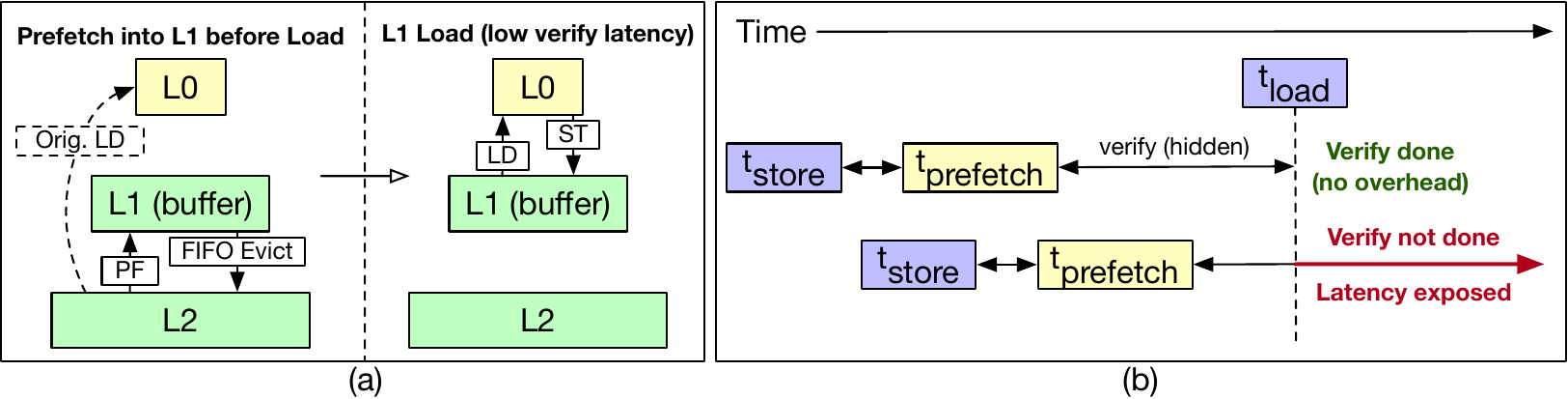}
        \caption{(\textbf{a})~When Trident-P encounters a load that accesses a program
                                qubit in the L2, Trident-P generates a prefetch instruction
                                that moves the program qubit into the L1.
                                To make space for the program qubit, Trident-P evicts the
                                oldest resident of the L1 and stores it into the L2.
                                This converts the L2 load into an L1 load with little
                                to no verification latency.
                  (\textbf{b})~The L2 must verify a prefetch, so an L1 load will incur
                                a verification overhead that depends on when the prefetch
                                occurs ($t_\mathrm{prefetch}$) relative to the load ($t_\mathrm{load}$).
                                Evicting the oldest stored program qubit (smallest $t_\mathrm{store}$)
                                maximizes the time between the prefetch ($t_\mathrm{prefetch}$) and the corresponding
                                load ($t_\mathrm{load}$).
                                Consequently, the L1 load incurs minimal verification overhead.
                                 }
        \label{fig:trident_p_overview}
    \end{figure*}
    
    While the speedups from both versions of Trident-D are substantial,
    there still remains a significant performance gap between Trident-D
    and an idealized system.
    Figure~\ref{fig:lru_opt_speedup} (bottom)  further shows the L1 hit rates,
    or proportion of loads that access the L1 memory, for Trident-D
    with LRU and Trident-D with OPT.
    Trident-D with LRU has a hit rate of 66.7\% and Trident-D with
    OPT has a hit rate of 78.7\%.
    For workloads like \texttt{chromium\_q\_sel}, both versions of
    Trident-D do not have high hit rates, and so there is a significant
    gap in performance between Trident-D and ideal.
    As we cannot do better than OPT in terms of cache hit rate, we must
    look towards an alternative implementation of Trident.

\section{Trident-P: Trident with Prefetching}
    Trident-D provides limited benefit on memory-intensive workloads
    with low temporal locality.
    Improving performance on these workloads requires moving toward
    a version of Trident whose performance is independent of the
    workload's temporal locality pattern.
    To this end, we propose \textit{Trident-P (Prefetch-Driven Trident)}
    which operates the memory hierarchy as a prefetch-driven hierarchy.
    Figure~\ref{fig:trident_p_overview} presents how Trident-P reduces
    verification latency.
    While a demand-driven hierarchy (Trident-D) installs the most frequently
    accessed program qubits in the L1, a prefetch-driven hierarchy moves
    program qubits in the L2 into the L1 before their next load.
    We refer to this as a \textit{prefetch}.
    
    Like Trident-D, Trident-P is also a compile-time solution that
    modifies a program already compiled to a two-level memory hierarchy.
    During compilation, Trident-P tracks the contents of the L1.
    Qubits can only enter the L1 if they are either (1)~stored from the L0,
    or (2)~prefetched from the L2.
    Trident-P does not modify loads that access the L1 memory as such loads
    have no verification latency.
    Thus, Trident-P focuses specifically on issuing prefetches to avoid L2
    loads.
    The remainder of this section discusses the technical details of prefetching
    and design of Trident-P.

    \subsection{Prefetch Verification Overheads}
        Prefetching a program qubit from the L2 requires verification.
        However, loading the program qubit from the L1 after the prefetch
        does not incur any additional verification overhead.
        Thus, an L1 load incurs some verification overhead dependent on
        when the prefetch executes relative to the load.
        If $t_\mathrm{load} - t_\mathrm{prefetch}$ is larger
        than the L2 verification latency (178 cycles), then the L1 load
        incurs no additional verification overhead as the L2 will verify
        the prefetch before the load occurs.
        Otherwise, the L1 load will have a verification overhead of
        $178 - (t_\mathrm{load} - t_\mathrm{prefetch})$ cycles.

    \subsection{Trident-P: Orchestrating Prefetches}
        Trident-P has two goals: eliminate L2 loads via prefetching, and ensure
        that the verification overhead of these prefetches is as low as possible.
        When Trident-P encounters an L2 load, Trident-P attempts to insert a prefetch
        instruction that installs the target program qubit into the L1.
        To install the target qubit into the L1, Trident-P must first allocate space
        by evicting a qubit currently residing within the L1.

        There are only two ways a program qubit can enter the L1: (1)~it was stored
        from the L0, or (2)~it was prefetched from the L2.
        We refer to the latter as \textit{inflight prefetches}.
        Trident-P will never evict an inflight prefetch as doing so degrades performance, 
        so it must choose a victim to evict among the qubits
        stored from the L0.
        However, the choice of victim can greatly affect the timeliness of a prefetch.
        Figure~\ref{fig:trident_p_overview}(b) presents how Trident-P selects
        which qubit to evict.
        Suppose an eviction candidate was installed in the L1 at time $t_\mathrm{store}$. 
        Then, $t_\mathrm{prefetch} > t_\mathrm{store}$ as the prefetch can only begin after the
        eviction candidate is stored into the L1.
        In order to maximize the difference between $t_\mathrm{load}$ and $t_\mathrm{prefetch}$,
        Trident-P chooses the eviction candidate with the smallest
        $t_\mathrm{store}$, or the \textit{oldest eviction candidate}.
        This is equivalent to FIFO eviction.

        Once Trident-P selects a program qubit to evict, it will generate a prefetch
        instruction which installs the target qubit into the L1 and installs the
        evicted qubit into the L2.
        Trident-P inserts the prefetch directly after the instruction that stored
        the victim qubit into the L1.
        This ensures that the prefetch occurs as early as possible.
        Trident-P may also not have any program qubit to evict, which occurs
        if the L1 is saturated with inflight prefetches.
        Then, Trident-P does not insert a prefetch and leaves the qubit to be loaded from L2.
        We note that this is incredibly rare as the L1 can buffer up to eight
        inflight prefetches, and in our evaluations, Trident-P rarely fails
        to generate a prefetch.

\ignore{
        \subsubsection{Preserving L1 Loads}
            A program qubit stored to the L1 from the L0 may be the target of a load
            instruction.
            Trident-P prevents this qubit from being evicted by marking it as ``safe,"
            as evicting the qubit will convert an L1 load into an L2 load.
            Thus, Trident-P's L1 eviction policy only considers L0 stores that have not
            been marked as ``safe."

        \subsubsection{Using OPT instead of FIFO for Eviction}
            As using FIFO eviction theoretically maximizes the difference between
            $t_{load}$ and $t_{prefetch}$, Trident-P uses FIFO eviction by default.
            Our experiments also corroborate our analysis.
            We find that Trident-P with OPT performs worse than Trident-P with FIFO
            as the mean load verification latency with OPT is larger.
}

\ignore{
    \subsection{Trident-P: Design}
        During compilation, Trident-P tracks the contents of the L1.
        When Trident-P encounters a load, it checks if the program qubit
        resides in the L1.
        If so, then Trident-P marks program qubit as ``safe" so it
        cannot be evicted.
        Otherwise, Trident-P attempts to prefetch the program qubit from the L2
        and place it into the L1.
        
        To make space for the target program qubit, 
        Trident-P must evict a qubit currently in the L1.
        Trident-P maintains the invariant that a program qubit can only reside in the
        L1 if (1)~it was stored from the L0, or (2)~it is an inflight prefetch from the L2.
        Evicting an inflight prefetch is detrimental, so Trident-P only evicts
        qubits stored from the L0.
        If a qubit is marked as ``safe," then it is exempt from eviction.
        Trident-P gathers all potential eviction candidates from the L1
        and chooses the oldest program qubit (FIFO eviction).
        Finally, Trident-P modifies the input program by adding a 
        prefetch instruction that installs the target qubit and
        evicted qubit into the L1 and L2, respectively.
        Note that if there are no eviction candidates available,
        then Trident-P does not prefetch.

    \subsection{Trident-P: Coverage and Timeliness}
        Trident-P's performance depends on the \textit{coverage} and
        \textit{timeliness} of its decisions.
        Coverage is the fraction of L2 loads for which a prefetch was issued.
        Timeliness is whether a prefetch was installed sufficiently well
        in advance to hide its L2 verification latency.
        We discuss the coverage and timeliness of Trident-P's decisions.
        For simplicity, we assume that no L0 stores are marked
        as ``safe" as this occurs rarely.

        \subsubsection{Coverage}
            Trident-P always generates a prefetch for an L2 load if
            there is space in the L1 for another inflight prefetch.
            The only situation where PROMPTLY cannot generate
            a prefetch is when the L1 is already saturated with
            inflight prefetches.
            As Trident-P maximizes the number of inflight prefetches,
            it maximizes coverage: it is impossible to increase
            coverage further.
            Nevertheless, the capacity of the L1 limits Trident-P's
            coverage as the L1's capacity bounds the number of
            concurrent inflight prefetches.
            Despite this limitation, we do not observe any case
            where Trident-P achieves less than 100\% coverage in
            our studies.
        
        \subsubsection{Timeliness}
            We numerically define the timeliness of a prefetch
            as in Equation~\eqref{eqn:timeliness}.
            $t_{load}$ is the cycle when the FTQC loads the target program
            qubit into the L0.
            $t_{prefetch}$ is the cycle when Trident-P prefetches and 
            installs the target program qubit into the L1.
            A larger $Timeliness$ is better and corresponds to
            a prefetch that hides a larger fraction of the L2's verification latency.
            \begin{equation}
                \label{eqn:timeliness}
                Timeliness = t_{load} - t_{prefetch}
            \end{equation}

            As $t_{load}$ is largely determined by the program, Trident-P
            can only affect $t_{prefetch}$.
            We explain how Trident-P maximizes $Timeliness$ in its prefetch
            decisions.
            Recall that when Trident-P installs a prefetch, it will evict
            a program qubit resident in the L1.
            Given Trident-P's invariant, this program qubit will have been 
            stored from the L0.
            If the time of the store is $t_{store}$, then
            $t_{prefetch} > t_{store}$.
            Furthermore, as Trident-P always evicts the oldest program qubit
            from the L1 via FIFO eviction, it minimizes $t_{store}$ and thus
            $t_{prefetch}$.
            Consequently, Trident-P indirectly maximizes $Timeliness$ for
            a prefetch.
}


\ignore{

    \subsubsection{Coverage}
        PROMPTLY always generates a prefetch for a load if there is
        space in the L1 memory for another inflight prefetch.
        The only situation where PROMPTLY cannot generate a prefetch is
        when it has already saturated the L1's capacity with inflight
        prefetches.
        So, PROMPTLY achieves the theoretical maximum coverage.
        Nevertheless, PROMPTLY's coverage is limited by the capacity of
        the L1, as the maximum number of concurrent inflight prefetches
        is bounded by the L1's capacity.
        Despite this limitation, we do not observe any case where PROMPTLY
        achieves less than 100\% coverage in our studies.

    \subsubsection{Timeliness}
        We numerically define the timeliness of a prefetch as described
        in Equation~\eqref{eqn:timeliness}.
        As shown, timeliness is the difference between the time the FTQC
        initiates a load ($t_{load}$) and the time a prefetch begins ($t_{prefetch}$).
        A large timeliness value is better and corresponds to a prefetch
        that hides a larger fraction of the L2's verification latency.
        As $t_{load}$ is fixed, PROMPTLY can only affect $t_{prefetch}$. 
        \begin{equation}
            \label{eqn:timeliness}
            Timeliness = t_{load} - t_{prefetch}
        \end{equation}
        Recall that during prefetch installation, PROMPTLY will evict
        some program qubit in the L1 to allocate space for the prefetched
        program qubit.
        Given PROMPTLY's invariant, this program qubit was originally stored
        from the compute subsystem and into the L1.
        If the time of this store is $t_{store}$, then $t_{prefetch} > t_{store}$.
        Furthermore, as PROMPTLY evicts the oldest such qubit, PROMPTLY minimizes
        $t_{store}$, thus maximizing the difference between $t_{load}$ 
        and $t_{prefetch}$.
        Thus, PROMPTLY's FIFO eviction decisions indirectly maximize timeliness.
        However, unlike traditional prefetchers, PROMPTLY does not avoid
        untimely prefetches.
        Note that even untimely prefetches still hide some of the L2's verification
        latency, even if the amount hidden is rather small.
        As untimely prefetches do not degrade performance, PROMPTLY does not
        explicitly avoid scheduling untimely prefetches.

\ignore{
    Both caching and prefetching have different tradeoffs.
    While a compiler-managed cache can make optimal replacement decisions
    as the memory access stream is known at compile-time, its performance
    is entirely determined by whether the memory access stream has high
    temporal locality.
    If the memory access stream has poor temporal locality, then
    a compiler-managed cache will provide poor performance as the program
    will likely thrash the cache.
    In contrast, prefetching into a latency hiding buffer is largely independent
    of temporal locality.
    Instead, the performance of a prefetcher is largely determined by whether
    its prefetches are \textit{timely}.
    For a prefetch to be timely, its verification must complete before the 
    corresponding load.
    If a prefetch is late, then its verification latency will stall the FTQC.
    Concurrently, a prefetcher must also provide timely prefetches to a
    majority of loads in the program: that is, it must have high \textit{coverage}.
}

\ignore{

\subsection{Architecture}
Prior work on large-scale surface code architectures typically assumes a two-level compute--memory hierarchy with an
unyoked surface code compute region backed by 2D yoked surface code storage. In this work, we augment this baseline
architecture  with a small 1D yoked surface code block, as shown in Figure \ref{fig:arch}. The resulting design forms a
three-level hierarchy with distinct roles. The \textbf{compute} region, composed of unyoked (0D) surface code patches,
holds qubits that are actively participating in computation. All program operations are executed in this region, so any
qubit residing in memory must first be loaded into compute before it can be operated on. The \textbf{backing memory},
composed of 2D yoked surface-code blocks, provides dense storage for qubits that are idle but must remain stored
reliably. Between these two, we introduce an \textbf{intermediate buffer} built from a 1D yoked surface code block. To
preserve the spatial efficiency of the overall memory system, this buffer is kept small, with the bulk of qubits
continuing to reside in 2D memory. The purpose of the intermediate region is to act as a low-capacity, low-latency
buffer from which qubits can be loaded much more quickly than from 2D memory.

\begin{figure}[h]
    \centering
    \includegraphics[width=0.75\linewidth]{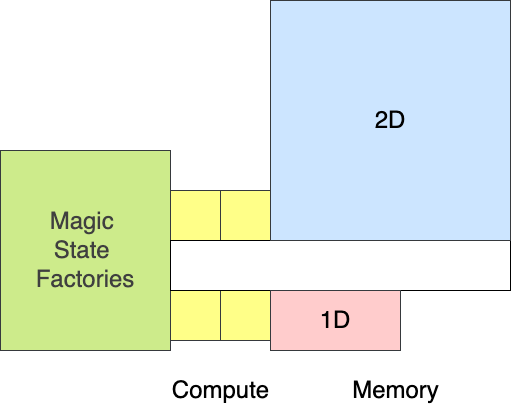}
    \caption{Three-level architecture consisting of 0D compute, 1D intermediate buffer, and 2D memory.}
    \label{fig:arch}
\end{figure}

\subsection{Advantages of the 1D Block as an Intermediate Buffer}

The key advantage of a 1D block is its short yoke-check sequence. Logical information in a 1D block is protected by only two yokes, both of which can be measured using two lattice surgery completed in 6 cycles. In contrast, a standard 2D block using the $[[256, 194, 4]]$ outer code requires 62 lattice surgery operations to check all of its yokes, which must be performed serially over 177 cycles. Since a qubit loaded from a yoked block is fully resolved only after the source block completes its next yoke-check sequence, a qubit loaded from 1D becomes ready for computations much sooner than one loaded from 2D. In fact, the two 1D yoke checks can be interleaved with the load operation with only 2 additional cycles, so a qubit loaded from 1D is ready for computation immediately after the load completes. A qubit can be loaded from or stored into a 1D block using a sequence of operations similar to that of a 2D block, so we model the same memory latency for 1D and 2D accesses. The notable difference between 1D and 2D accesses is therefore the readiness latency, which is much larger for qubits loaded from 2D.

\subsection{Data Movement}

Due to the no-cloning theorem, quantum states cannot be copied. Every load or store destroys the information at its
source location. A quantum memory system therefore forms an \emph{exclusive hierarchy}, in which each logical qubit
resides in exactly one region at a time. In our architecture, each region operates at capacity, so bringing a qubit into
compute requires simultaneously moving another qubit out to memory to free space. Accordingly, load and store operations
are coupled by design.

A three-level hierarchy introduces new data movement patterns on top of those in a two-level system. In addition to
transfers between compute and either storage region, it also permits movement across all three levels or between storage
levels. For this architecture, we define two core primitives that are sufficient to express the policies studied in this
work.

\subsubsection{Three-way cyclic shift}

Suppose we wish to load a qubit from backing memory into compute, while storing the compute-resident qubit into the
intermediate buffer because it is expected to be reused. To make room in the buffer, a resident qubit must be evicted
from the buffer back to backing memory. We denote this three-way movement by
\[
\mathrm{CYCLIC\_SHIFT}(ld,st,ev),
\]
where \(ld\) is loaded from backing memory into compute, \(st\) is stored from compute into the buffer, and \(ev\) is
evicted from the buffer back to memory.

Buffering the qubit allows later accesses to avoid the long readiness delay associated with a fresh load from backing
memory.

\subsubsection{Two-way intermediate stage}

We also allow movement between the buffer and backing memory without involving compute. A qubit in memory can be staged
into the buffer before it is demanded by compute. To make room in the buffer, a resident qubit must be evicted from the
buffer back to backing memory. We denote this two-way movement by
\[
\mathrm{STAGE}(sg,ev),
\]
where \(sg\) is staged from backing memory into the buffer and \(ev\) is evicted from the buffer to memory. 

The benefit of moving a qubit from memory into the intermediate buffer depends on when it is staged, and choosing this timing is a key challenge in using this primitive
effectively.

Given the three-level architecture and these primitives, the central policy question is how qubits should be placed into and evicted from the
intermediate buffer over time. 

\subsection{Cache-Based Reactive Placement}

A natural first approach is to treat the buffer as a small, low-latency \emph{cache} for
the backing memory. The goal is to retain qubits that are likely to be reused, so that subsequent accesses can be served
from the buffer with no readiness latency. Cache-based placement presents a simple solution because qubits written into
the buffer from compute are typically already fully resolved, avoiding the more delicate timing considerations of
staging qubits from 2D memory.

Under this placement policy, qubits are initially loaded from backing memory into compute. After completing their
operations, they are written into the intermediate buffer. If the buffer is full, one resident qubit is selected for
eviction and written back to backing memory. If a qubit still resides in the buffer when it is next needed, the access
is a \emph{cache hit} and can be served from the buffer with no readiness latency. Otherwise, the access is a
\emph{cache miss}. The qubit must be loaded again from backing memory and incur the long readiness latency of the 2D
block. It is therefore desirable to maximize the fraction of loads served from the buffer. So, the primary performance
metric for cache-based placement is the \emph{hit rate}.

In terms of implementation, cache hits can be served with a simple coupled load-store. Since the qubit being loaded resides in the buffer, loading it vacates a buffer slot, and the stored qubit is simply placed into that slot. For cache misses, the qubit must be brought in from backing memory while the compute-resident qubit is written into the buffer. Since the buffer operates at capacity, one resident qubit in the buffer must be evicted to backing memory. This can be accomplished with the $\mathrm{CYCLIC\_SHIFT}$ primitive. The remaining question is which resident qubit should be evicted when a new qubit needs a slot, which is guided by the cache replacement policy. 

We consider two standard policies from classical memory systems.

\begin{enumerate}
    \item \textbf{Least Recently Used (LRU)}: LRU uses recency as a proxy for future reuse. The eviction victim is the
    qubit whose most recent access lies furthest in the past.
    \item \textbf{Belady's Optimal (OPT)}: OPT uses future knowledge of the program trace. The eviction victim is the
    qubit whose next use lies farthest in the future.
\end{enumerate}

\paragraph{Example}
Consider the example in Figure~{\color{red} XX}. Both the compute region and the intermediate buffer have capacity 2.
Initially, compute holds \(q_0\) and \(q_1\), the buffer holds \(q_2\) and \(q_3\), and \(q_4\) and \(q_6\) reside in
backing memory. Instruction A executes immediately because both of its operands are already in compute. Instruction B
requires \(q_2\), which is currently in the buffer. Loading \(q_2\) into compute vacates a buffer slot, so \(q_0\) is
stored into that slot using a coupled load-store. Because a load from the buffer has no readiness delay, instruction B
can execute immediately after the transfer. Instruction C requires \(q_4\), which must be loaded from backing memory,
while the compute-resident qubit \(q_1\) is stored into the buffer. Since the buffer is full, either \(q_0\) or \(q_3\)
must be evicted. Because \(q_3\) is  unused, both LRU and OPT evict \(q_3\), implemented with a simultaneous
\(\mathrm{CYCLIC\_SHIFT}(q_4, q_1, q_3)\). Moreover, because \(\mathrm{CRZ}\) is non-Clifford operation for arbitrary
angles, instruction C must wait until the loaded state of \(q_4\) is resolved before it can execute. Instruction D next
reloads \(q_1\) from the buffer, so it can again execute immediately after the transfer. Instruction E then requires
\(q_6\) from backing memory while storing \(q_1\) back into the buffer. At this point, the replacement policies diverge:
among the buffer residents, \(q_4\) is the most recently used, whereas \(q_0\) is the qubit whose next use occurs
soonest. Accordingly, LRU evicts \(q_0\), while OPT evicts \(q_4\). As a result, instruction F, which requires \(q_0\),
is a cache miss under LRU but a cache hit under OPT. The resulting hit rates are \(40\%\) for LRU and \(60\%\) for OPT.
Note also that the accesses to \(q_4\) and \(q_6\) are compulsory misses, since neither qubit had been accessed
previously. No reactive cache-based replacement policy could have placed them in the buffer ahead of demand.

Cache-based replacement exploits temporal locality. When the same qubits are brought into compute repeatedly, the hit rate is high. Figure \ref{fig:cache} shows the hit rates across our benchmarks. Cache hits can substantially improve performance over the baseline architecture by allowing nearby accesses to be served from the buffer instead of backing memory. Additionally, since Belady's algorithm is designed to provide the optimal hit rate in a replacement policy, it outperforms LRU replacement. Thus, in our studies on caching, we primarily use Belady's replacement.

\begin{figure}
    \centering
    \includegraphics[width=1\linewidth]{figures/cache_lru_opt.png}
    \caption{Hit rates of LRU and OPT replacement}
    \label{fig:cache}
\end{figure}

However, we observe that the hit rate is not sufficiently high across all benchmarks. Benchmarks that access a wide range of qubits in consecutive accesses with low hit rates that do not adequately reduce the mean readiness latency of a memory access. This highlights the main limitation of caching: cache-based policies are inherently reactive. The buffer is populated only after a qubit has already been used in compute. It does nothing for the first future access to a qubit that is still sitting in 2D memory. For certain programs, this can sharply reduce the benefit of caching and leave a substantial gap to ideal performance. This motivates proactive staging: qubits must be moved into the buffer before they are first demanded in compute, rather than only after they have already been used once.

\subsection{Prefetching-Based Proactive Placement}

A proactive placement policy would aim to stage a qubit from backing memory to the intermediate buffer ahead of its
load, so that the load can be serviced from the buffer. Thus, qubits can be \emph{prefetched} to the buffer in advance.
If this is done sufficiently early, the long readiness delay of loading from backing memory can elapse while the qubit
resides in the buffer, moving it off the critical path of execution.

Under this policy, a qubit can be loaded into compute from the buffer or from backing memory using a coupled load-store. The stored qubit replaces the loaded qubit in its respective storage. Additionally, a qubit that needs to be loaded from backing memory can be staged in the buffer at some earlier point
in the program using the $\mathrm{STAGE}$ primitive. Once the buffer is full, the stage must simultaneously evict a qubit from the buffer to backing memory.  Buffering a qubit in 1D buffering alone does not eliminate the uncertainty inherited from a fresh load out of 2D. If a qubit is
staged from memory into the buffer, the state in the buffer is not fully resolved until the source block completes its next
yoke-check sequence. If the qubit is loaded into compute before the sequence completes, it must still wait to be resolved and stall computation.

For this reason, the hit rate alone is not a good enough metric to quantify the performance of a prefetcher. Naively inserting prefetches prior to every memory load can produce a very high hit rate. If a qubit is prefetched into the buffer and then immediately loaded into compute, it is
unlikely that it absorbed any of the 2D block readiness latency, stalling its operations while having occupied limited
buffer capacity that could have been used by a more valuable prefetch. A good prefetcher must also ensure \emph{timeliness}, so that the prefetch is inserted far enough in advance of its load to absorb a significant portion of the verification latency. We collect the average distance between an inserted prefetch and its corresponding load and quantify a compile-time estimate of verification latency, which can be used to compare the performance impact of different policies.

So, prefetching must decide 1) which loads should be prefetched, 2) when in the program the prefetch is inserted, and 3) 
which buffer qubit to evict. To decide this, we track prior $\mathrm{MSWAP}$ operations that stored qubits into the 1D
buffer. For a target load $\mathrm{MSWAP}(l,s)$, these prior stores define candidate points after which a prefetch could
be inserted and candidate qubits that could be evicted. Among the tracked candidates, we retain only those that where the target $\mathrm{MSWAP}(l,s)$ indirectly depends on them in the instruction DAG.
The dependency condition ensures that inserting the prefetch does not alter inter-qubit dependencies
in the DAG.

Among the remaining feasible candidates, the scheduler selects the eviction target using the same policies as in the
cache-based baseline. Under LRU, the selected victim is the least recently used feasible 1D resident. Under Belady, the
selected victim is the feasible 1D resident whose next use lies furthest in the future. In this sense, prefetching extends
cache-based placement. A prefetch policy may choose to retain useful resident qubits by declining to evict them, so it
still captures the temporal-locality benefits of caching. But unlike reactive caching, it can also reduce the latency of
first future accesses to qubits that are currently in backing memroy. This suggests that prefetching may offer an
additional advantage beyond what reactive caching alone can provide.

\paragraph{LRU vs OPT for prefetching} While prefetching a qubit $p$ from memory, our goal is to maximize the verification latency hidden by the buffer. This is achieved when the prefetch is inserted far in advance of its load. When choosing among a list of candidates to replace in the buffer, it is natural to choose the one stored in the buffer furthest in the past, say $e$. A $\mathrm{STAGE}(p,e)$ can be inserted immediately after $e$ is stored in the buffer, allowing the largest time to ellapse before $p$ is ultimately loaded to compute.  Such a choice is made by an LRU eviction policy, which evicts the candidate stored in buffer earliest in the past. On the other hand, OPT maximizes the number of cache hits, which reduces the number of prefetches that need to be issued. Cache hits are desirable since they always incur a zero verification latency. However, this comes at the cost of prefetch timeliness, since the prefetch would be inserted closer to the load than with LRU. In fact, accounting for the smaller distances between a prefetch inserted by OPT to its load, even with a larger hit rate the expected verification latency of OPT is higher than LRU, as shown in Figure \ref{fig:prefetch}.

There is another disadvantage of using OPT: a qubit can occupy a slot in the buffer for long periods of time, potentially thousands of cycles, in between successive accesses. This is suboptimal even if it increases the number of cache hits. This buffer slot could have instead been used to prefetch qubits for other memory accesses. Say the verification latency of a load is 200 cycles. 1000 cycles can be used to hide the latency of 5 different accesses, instead of being used up by a single qubit in OPT. Thus, in practice, LRU eviction is more performance than OPT eviction for prefetching, so unless otherwise noted we report results for prefetching with LRU eviction.

\ignore{
The threshold $\tau$ controls the tradeoff between hit rate and readiness latency. A larger $\tau$ makes prefetching
more conservative because fewer qubits are prefetched, lowering hit rate, but those that are prefetched are more likely
to be fully ready when loaded into compute. A smaller $\tau$ admits more prefetches, increasing hit rate, but also
increases the likelihood that a qubit reaches 1D too late to fully hide the underlying 2D delay. The best value of
$\tau$ balances these competing effects.
}

\begin{figure}[h]
    \centering
    \includegraphics[width=1\linewidth]{prefetch_verify_expect.png}
    \caption{Expected time to verify for LRU and OPT replacement}
    \label{fig:prefetch}
\end{figure}
}}

\begin{figure*}
    \centering
    \includegraphics[width=\textwidth]{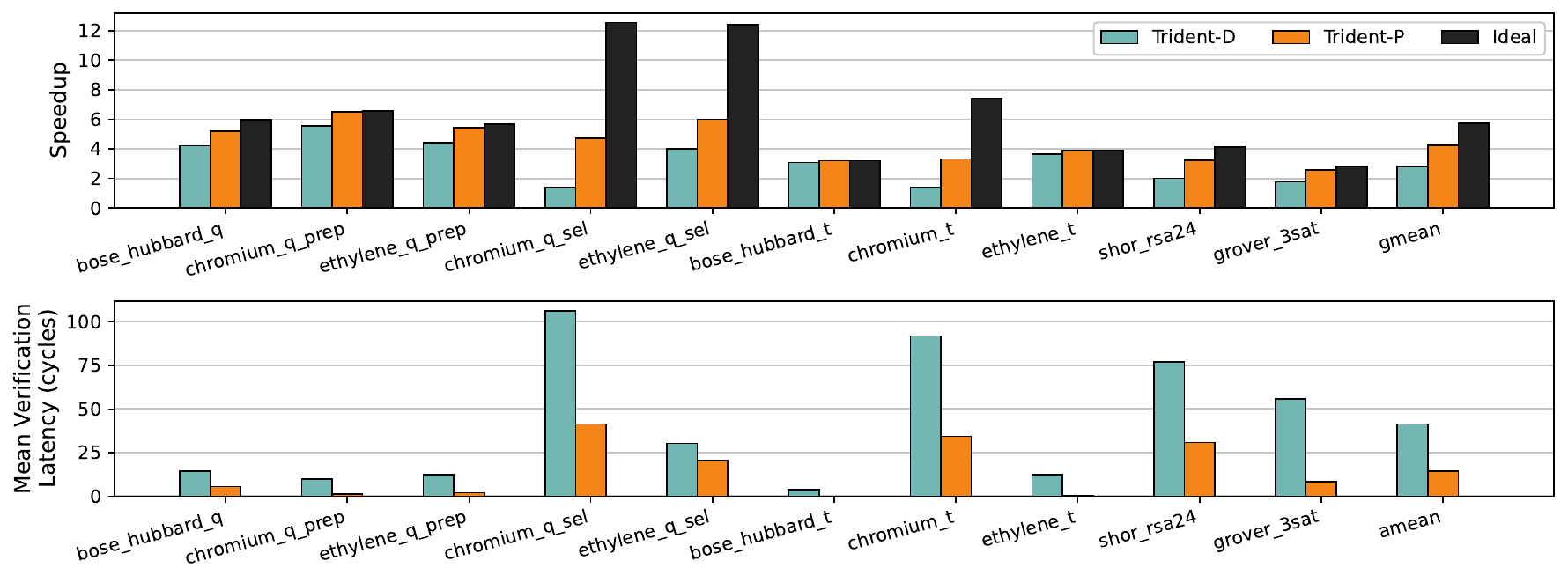}
    \caption{\textbf{(Top)} Speedups of Trident-D cache replacement, Trident-P, and an idealized system with zero verification latency relative to
            the baseline system.
            Trident-D, Trident-P and the idealized system achieve speedups of $2.8\times$, $4.2\times$, and $5.7\times$, respectively. \textbf{(Bottom)} The average load verification latency with Trident-D and Trident-P.
                        Trident-D and Trident-P achieve mean load verification latencies of
                        41.3 and 14.3 cycles, respectively.}
    \label{fig:performance}
\end{figure*}

\section{Results}


    \subsection{Impact on Instruction Throughput}
        Figure~\ref{fig:performance} (top) compares the speedups of Trident-D, Trident-P, and the idealized system
        over the baseline system with a two-level memory hierarchy.
        Trident-D, Trident-P, and the idealized system provide average speedups of $2.8\times$, $4.2\times$, and $5.7\times$,
        respectively.
        Trident-P outperforms Trident-D across all workloads.
        The greatest gap between Trident-P and Trident-D is for memory-intensive workloads such as
        \texttt{chromium\_q\_sel} that exhibit poor temporal locality and thrash Trident-D's cache.
        There is a significant gap between Trident-P and ideal for these memory-intensive workloads.
        
\begin{figure}[!htb]
            \centering
            \includegraphics[width=1\linewidth]{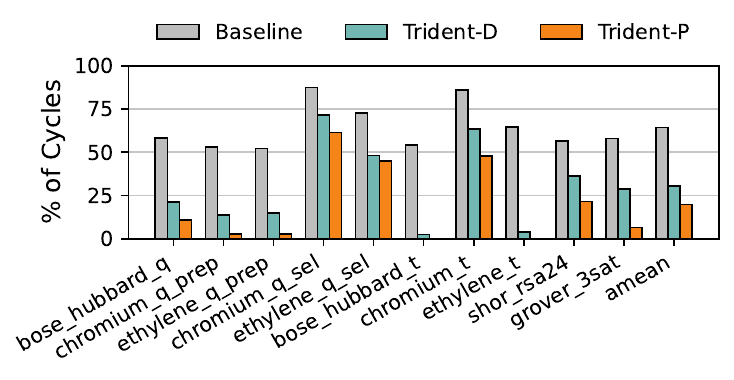}
            \caption{Percentage of cycles stalled due to verification ($64\%$ in baseline, $30\%$ with Trident-D, and $20\%$ with Trident-P).}
            \label{fig:verification}
        \end{figure}

        Figure~\ref{fig:performance} (bottom) shows the average load verification latency
        for Trident-D and Trident-P (baseline verification latency is 178 cycles).
        Trident-D and Trident-P reduce the mean verification latency to 41.3 and 14.3 cycles, respectively. Trident's reduction in verification latency also reduces 
        verification-related stalls, as shown in Figure~\ref{fig:verification}.
        Trident-D and Trident-P reduce load verification stalls from 64\% of cycles to 30\% and 20\%, respectively.

\subsection{Trident-P: Prefetch Coverage and Timeliness}

\begin{figure}[!htb]
    \centering
    \includegraphics[width=1\linewidth]{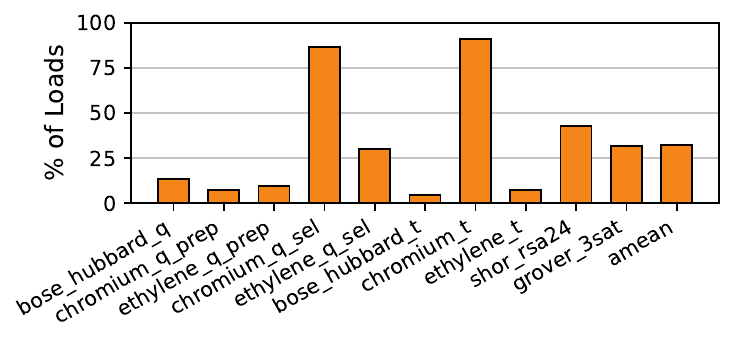}
    \caption{Prefetch Coverage for Trident-P. Trident-P inserts prefetches for 32.5\% of loads.}
    \label{fig:coverage}
\end{figure}

Figure \ref{fig:coverage} shows that Trident-P's lower latency is driven by prefetches. For benchmarks such as \texttt{chromium\_q\_prep} and \texttt{ethylene\_t}, Trident-D achieves high L1 hit rates, leaving few L2 loads for Trident-P to prefetch. On the other hand, benchmarks such as \texttt{chromium\_q\_sel} and \texttt{chromium\_t} have lower L1 hit rates and therefore more L2 loads, enabling Trident-P to achieve higher prefetch coverage. Across benchmarks, Trident-P prefetches 32.5\% of loads on average. 

\begin{figure}[!htb]
    \centering
    \includegraphics[width=1\linewidth]{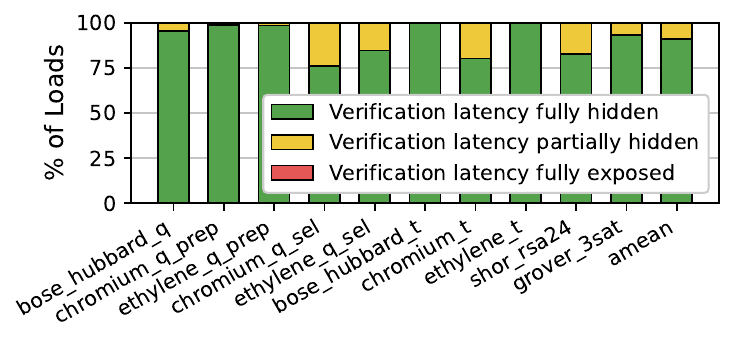}
    \caption{Verification latency hiding in Trident-P. Latency is fully hidden for 91.0\%, partially hidden for 9.0\%, and fully exposed for $<0.01\%$ of loads.}
    \label{fig:breakdown}
\end{figure}

Further, Figure \ref{fig:breakdown} provides evidence that these prefetches are timely.  Trident-P fully hides verification latency for 91.0\% of loads and partially hides it for another 9.0\%, improving over Trident-D's hit rate of 78.7\% (Figure~\ref{fig:lru_opt_speedup}). Verification latency remains fully exposed for fewer than 0.01\% of loads, indicating that Trident-P almost always issues prefetches early enough to hide at least part of the verification.

\subsection{Sensitivity to L1 Memory Capacity}

\begin{figure}[!htb]
    \centering
    \includegraphics[width=1\linewidth]{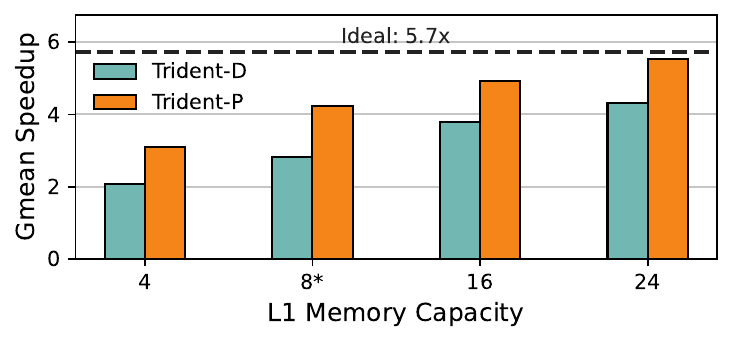}
  
    \caption{Speedups of Trident-D and Trident-P for L1 memory capacities of 4, 8, 16, and 24.
            Trident-D provides speedups of $2.1\times$, $2.8\times$, $3.8\times$, and $4.3\times$, respectively.
            Trident-P provides speedups of $3.1\times$, $4.2\times$, $4.9\times$, and $5.5\times$, respectively.}
    \label{fig:buffer_capacity}
\end{figure}

Figure~\ref{fig:buffer_capacity} compares the performance of Trident-D and Trident-P for L1 memory capacities
of 4, 8 (default), 16, and 24.
As shown, Trident-D provides speedups of $2.1\times$, $2.8\times$, $3.8\times$, and $4.3\times$, respectively.
Trident-P provides speedups of $3.1\times$, $4.2\times$, $4.9\times$, and $5.5\times$, respectively.
The performance of Trident-D and Trident-P improves almost linearly as the L1's capacity increases.
Furthermore, Trident-P at a smaller capacity (i.e., 8) outperforms Trident-D at the next largest capacity (i.e., 16).
Thus, Trident-P requires a lower physical qubit footprint to achieve a comparable speedup to Trident-D.

\subsection{Sensitivity: Compute Capacity Size}

\begin{figure}[!htb]
    \centering
    \includegraphics[width=1\linewidth]{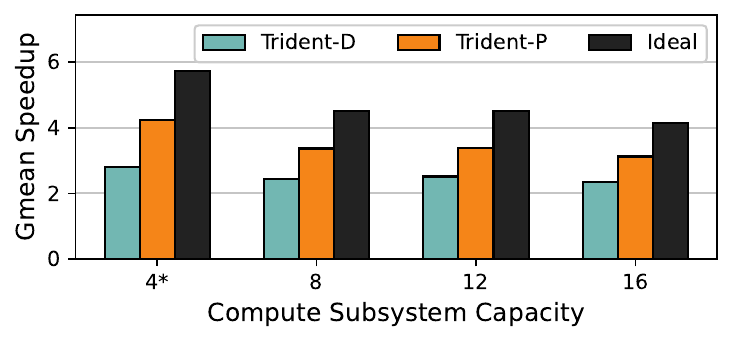}
    \caption{Speedups for Trident-D and Trident-P for different compute subsystem capacities.
            All speedups are normalized to the baseline system with an identically-sized compute subsystem.
            Trident-D provides speedups of $2.8\times$, $2.4\times$, $2.5\times$, and $2.3\times$, respectively.
            Trident-P provides speedups of $4.2\times$, $3.4\times$, $3.4\times$, and $3.1\times$, respectively.}
    \label{fig:compute_capacity}
\end{figure}

Increasing the capacity of the compute subsystem (L0) 
reduces the number of memory accesses required to execute
a quantum program and thus reduces memory-related overheads.
Hence, we study how Trident-D and Trident-P scale to FTQCs with larger compute subsystems.
Figure \ref{fig:compute_capacity} compares Trident-D and Trident-P for compute subsystem capacities
of 4 (default), 8, 12, and 16. 
The reported speedups for each capacity are normalized to the baseline with the same capacity.
Trident-D provides speedups of $2.8\times$, $2.4\times$, $2.5\times$, and $2.3\times$, respectively.
Trident-P provides speedups of $4.2\times$, $3.4\times$, $3.4\times$, and $3.1\times$, respectively.
Trident-D and Trident-P provide speedups even when memory-related overheads reduce, demonstrating
that they will scale to FTQC systems with larger compute subsystems.

\subsection{Sensitivity: 2D Block Size}
In this paper, we assume that each 2D YSC in the L2 memory uses an outer code of $[[256, 194, 4]]$ as this is typically used
in prior works~\cite{gidney2025_rsa_factorization, khattar2025_verifiable, babbush2026_ecc256}.
Depending on the number of program qubits in a quantum program, 
a larger outer code may be more desirable as the storage density
of a yoked surface code increases with the size of its outer code.
However, a larger outer code inherently has more yokes, so its verification latency is longer.
Equation~\eqref{eqn:verify_latency} shows how the verification latency scales with
the size of the outer code.
As shown, increasing the number of surface code qubits within a YSC, or the block size of the YSC,
by $4\times$ increases verification latency by $2\times$.
\begin{equation}
    \label{eqn:verify_latency}
    \mathrm{VerificationLatency} \propto 25 \times \sqrt{\mathrm{BlockSize}}
\end{equation}

Table~\ref{tab:block_size_sens} shows the speedups of Trident-D, Trident-P, and ideal (no verification latency)
for five different outer codes.
The five outer codes have 64, 144, 256, 400, and 576 constituent surface code qubits which are
encoded into 34, 98, 194, 322, and 482 logical qubits, respectively.
Trident-D provides speedups of $3.7\times$, $3.1\times$, $2.8\times$, $2.6\times$, and $2.4\times$, respectively.
Trident-P provides speedups of $4.6\times$, $4.3\times$, $4.2\times$, $4.0\times$, and $3.8\times$, respectively.
An idealized memory with no verification latency has a speedup of $5.7\times$ regardless of block size.
The speedups of Trident-D and Trident-P decrease for larger block sizes.
This is expected as a larger verification latency is harder to mitigate using the L1 memory.
However, the speedups of Trident-D and Trident-P both remain significant even for the largest
outer code evaluated.

\begin{table}[!htb]
    \begin{center}
    \caption{Speedups for Different 2D YSC Blocks}
    \label{tab:block_size_sens}
    \footnotesize
    \begin{tabular}{|c||c|c||c|c|c|}
        \hline
        Block Size & Capacity & V.L. & Trident-D & Trident-P & Ideal\\
        \hline
        \hline
        64 & 34 & 90 & $3.7\times$ & $4.6\times$ & $5.7\times$\\
        144 & 98 & 134 & $3.1\times$ & $4.3\times$ & $5.7\times$ \\
        256 & 194* & 178 & $2.8\times$ & $4.2\times$  & $5.7\times$\\
        400 & 322 & 222 & $2.6\times$ & $4.0\times$  & $5.7\times$\\
        576 & 482 & 266 & $2.4\times$ & $3.8\times$ & $5.7\times$ \\
        \hline
    \end{tabular}
    \end{center}
\end{table}

\ignore{

\begin{figure}[h]
    \centering
    \includegraphics[width=1\linewidth]{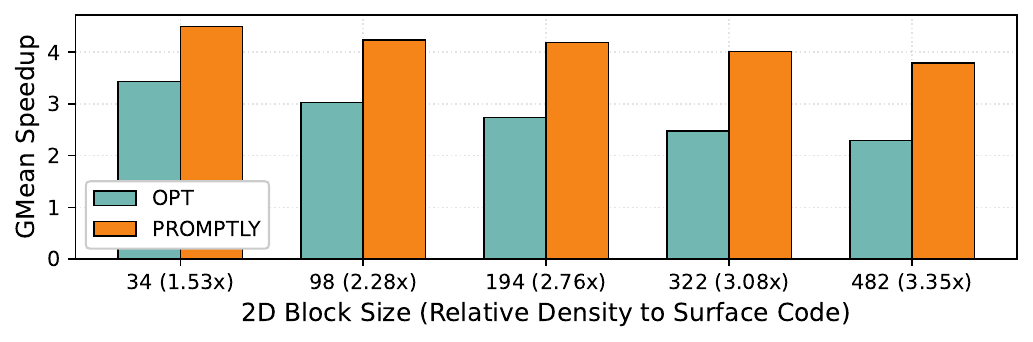}
    \caption{Gmean speedups for 2D block sizes: $[[64, 34, 4]]$, $[[144, 98, 4]]$, $[[256, 194, 4]]$, $[[400, 322, 4]]$, $[[576, 482, 4]]$. Performance degradation for Trident-P is slower than for Trident-D.}
    \label{fig:block_size}
\end{figure}
}
\subsection{Impact on Application Fidelity}

\textit{Application fidelity} is largely determined by the memory-errors incurred during program execution. Since Trident reduces the cycles required to execute a program, it improves fidelity over the baseline. As shown in Table \ref{tab:fidelity}, application fidelity tracks program speedup: it is lowest for the baseline at 92.9\%, higher for Trident-D at 97.4\%, and highest for Trident-P at 98.1\%. Note that the idealized system achieves 100\% application fidelity for every benchmark.

\begin{table}[!htb]
    \begin{center}
    \caption{Application Fidelities across Benchmarks}
    \label{tab:fidelity}
    \footnotesize
    \begin{tabular}{|c||c|c|c|}
        \hline
        Benchmark & Baseline & Trident-D & Trident-P \\
        \hline\hline
bose\_hubbard\_q & 90.2\% & 97.7\% & 98.1\% \\
chromium\_q\_prep & 97.5\% & 99.6\% & 99.6\% \\
ethylene\_q\_prep & 90.4\% & 97.9\% & 98.3\% \\
chromium\_q\_sel & 85.8\% & 89.9\% & 93.5\% \\
ethylene\_q\_sel & 75.9\% & 94.1\% & 95.3\% \\
bose\_hubbard\_t & 97.0\% & 99.0\% & 99.1\% \\
chromium\_t & 97.2\% & 98.2\% & 98.3\% \\
ethylene\_t & 98.5\% & 99.6\% & 99.7\% \\
shor\_rsa24 & 99.6\% & 99.8\% & 99.9\% \\
grover\_3sat & 96.8\% & 98.2\% & 98.8\% \\
\hline\hline
Average & 92.9\% & 97.4\% & 98.1\% \\

        \hline
    \end{tabular}
    \end{center}
\end{table}

\subsection{Impact on Physical Qubit Footprint}

In terms of physical qubit footprint, the baseline two-level memory hierarchy uses the fewest physical qubits but also has the worst performance.
On the other end of the spectrum, the idealized two-level memory hierarchy has the best performance but requires the most physical
qubits as it only uses 1D YSCs within its memory subsystem.

Table~\ref{tab:storageeff} shows the increase in physical qubit footprint of Trident
and the idealized FTQC relative to the baseline.
In our analysis, we consider the physical qubit footprint of program qubits
and associated routing space in the compute and memory subsystems (excluding magic state factories).
Furthermore, while we evaluated the \texttt{PREPARE} and \texttt{SELECT} subroutines
for \texttt{chromium\_q} and \texttt{ethylene\_q} separately, we report a single
value here for the entire workload to avoid inflating our results.
As shown, Trident has a $4.9\%$ higher physical qubit footprint relative to the baseline.
In contrast, the idealized FTQC has a $52.8\%$ higher physical qubit footprint.
Thus, Trident achieves 73.7\% of the performance of the idealized FTQC while requiring
far fewer physical qubits to implement.

\ignore{
Trident uses a small amount of 1D YSC to reduce the verification latency of 2D YSC.  While this reduces overall storage efficiency relative to a memory using only 2D YSC, the resulting overhead is modest. Table~\ref{tab:storageeff} compares the storage efficiency of three architectures: the baseline two-level hierarchy with 0D+2D YSCs, Trident's three-level hierarchy with 0D+1D+2D YSCs, and an idealized two-level hierarchy with 0D+1D YSCs and no verification latency. Storage efficiency is measured in terms of the number of physical qubits per logical qubit, normalized to a 0D surface code patch at the same effective code distance. The baseline is $2.37\times$ as dense as 0D surface code, while Trident remains $2.25\times$ as dense. By comparison, the idealized 0D+1D hierarchy is only $1.43\times$ as dense. Thus, Trident incurs only a $5\%$ increase in physical qubit cost relative to the baseline, while delivering $75\%$ of the performance benefit of an idealized memory system. 
}

\begin{table}[!htb]
    \begin{center}
    \caption{Increase in Footprint Relative to Baseline}
    \label{tab:storageeff}
    \footnotesize
    \begin{tabular}{|c|c||c|c|}
        \hline
        Benchmark & Qubits & Trident & Ideal (1D YSC)\\
        \hline
        \hline
        bose\_hubbard\_q & 2228 & $+0.8\%$ & $+92.7\%$\\
        chromium\_q & 167 & $+9.1\%$ & $+66.1\%$ \\
        ethylene\_q & 289 & $+4.8\%$ & $+47.0\%$\\
        bose\_hubbard\_t& 2188 & $+0.8\%$  & $+89.2\%$\\
        chromium\_t & 121 & $+9.2\%$ & $+22.1\%$ \\
        ethylene\_t & 241 & $+4.8\%$ & $+23.0\%$ \\
        shor\_rsa24 & 111 & $+9.1\%$ & $+12.9\%$ \\
        grover\_3sat & 1710 & $+1.1\%$ & $+96.8\%$ \\
        \hline
        \hline
        \multicolumn{2}{|c|}{gmean} & $+4.9\%$ & $+52.8\%$\\
        \hline
    \end{tabular}
    \end{center}
\end{table}

Trident can further reduce its physical qubit overhead by reusing an existing 1D YSC block as L1 (Appendix~\ref{appendix:fragmentation}).
\section{Related Work}

\subsection{Yoked Surface Codes}

Yoked surface codes (YSCs) were introduced by Gidney et al. as components of a surface-code memory hierarchy compatible with planar superconducting hardware \cite{gidney2023_yoked_sc}.  Since then, YSCs have primarily appeared in resource-estimation studies for large scale applications including 256-bit elliptic-curve discrete logarithm, RSA-2048 factoring, and optimal polynomial intersection  \cite{babbush2026_ecc256, gidney2025_rsa_factorization, khattar2025_verifiable}. These works treat YSCs as dense storage for a large number of qubits that are idle during computation, helping reduce the required hardware for such applications to under one million physical qubits. However, these studies do not examine how YSCs behave during execution: when data should be moved, whether the memory accesses are fault-tolerant, or how post-load verification latency impacts instruction throughput.
Indeed, the problem of load verification was raised by Gidney recently~\cite{gidney_stack_exchange_yoke_blocking}, but
recent works studying the yoked surface code have not accounted for the
impact of load verification latency.

Trident complements existing research on yoked surface codes by demonstrating that
the high load verification latency of the 2D yoked surface codes can be overcome by
expanding the traditional two-level memory hierarchy into a three-level memory hierarchy
with an intermediate memory tier containing a single 1D YSC block.
We propose two methods of managing the three-level memory hierarchy: Trident-D (demand-driven)
and Trident-P (prefetch-driven).

\subsection{Quantum Memory Hierarchies}

A growing body of work studies segregated compute--memory organizations for fault-tolerant quantum computing, 
but under different code and hardware modalities.
\textbf{LSQCA}~\cite{kobori2025_lsqca} uses surface codes in both the compute and memory subsystem.
Unlike the compute subsystem, which has ample routing space in their studies, the memory subsystem has
very limited routing space, resulting in high memory access latency.
LSQCA studies how memory management techniques can exploit temporal and spatial locality to improve
the average access latency to the memory subsystem. 
\textbf{HetEC}~\cite{stein2025_hetec} studies FTQCs with a surface code compute subsystem and 
a memory subsystem composed of Bivariate Bicycle codes~\cite{bravyi2023_gross_codes, yoder2025_tour_de_gross}.
HetEC finds that an FTQC with a surface code compute subsystem and QLDPC-based memory subsystem provides
comparable performance to FTQCs that exclusively use the surface code while using far fewer physical
qubits.
\textbf{qSIEVE}~\cite{viszlai2026qsieve} studies the same architecture for neutral atom systems.
qSIEVE also examines data movement challenges and optimizes the syndrome extraction schedule for
Bivariate Bicycle codes on neutral atom systems.

YSCs introduce unique challenges not present in QLDPC-based memory systems
because they are concatenated codes whereas high-rate QLDPC codes are implemented directly using physical qubits.
Consequently, the problem of load verification is unique to YSCs.
To the best of our knowledge, no similar problem exists for QLDPC-based memory systems.
As this paper studies the problem of reducing load verification overheads in a YSC
memory system, prior works like LSQCA, HetEC, and qSIEVE are orthogonal to this paper.

\subsection{Memory Systems Using QLDPC Codes}

Load/store architectures using high-rate QLDPC codes as a memory initially
came out of Cohen et al.'s work on code surgery~\cite{cohen2022_ldst}.
Following this work,
\textbf{HetEC} and \textbf{qSIEVE} study FTQCs whose
memory subsystem uses Bivariate Bicycle (BB) codes~\cite{stein2025_hetec, viszlai2026qsieve}.
\textbf{Tour de Gross} is a proposal from IBM that performs computation entirely using BB 
codes~\cite{yoder2025_tour_de_gross}.
The \textbf{Pinnacle} architecture from Iceberg Quantum proposes using QLDPC
codes for computation and memory~\cite{webster2026_pinnacle}.

While these architectures can be implemented on any hardware substrate,
there is great uncertainty about the viability of implementing 
high-rate QLDPC codes like BB codes on superconducting quantum 
processors~\cite{gent2025_ibm_loon_progress, wang2026_gross_code_demo}.
High-rate QLDPC codes require dense and non-planar connectivity,
which is hard to implement using superconducting qubits because of
crosstalk and frequency 
collisions~\cite{vittal2024_fpn, li2021_qsw_hw_codesign, murali2020_crosstalk}.
In contrast, yoked surface codes do not require any hardware changes as they
build upon the surface code which has been demonstrated several times on
real systems~\cite{krinner2022_distance3_demonstration, google2024_distance7_sc, google2023_d5_surface_code}.
Thus, there is less uncertainty about whether yoked surface codes can be realized
on real systems.

\ignore{
\section{Discussion}

We discuss alternate ways to handle uncertainty in the correctness of loads from a 2D YSC block. Throughout this work, we adopt the most conservative architectural choice of stalling computation until the source block checks all its yokes and the load is verified. An alternative approach would be to continue executing speculatively, assuming the load completed correctly, analogous to runahead execution in classical systems. If a later yoke check reveals that the load was faulty and a dependent non-Clifford operation has already been executed, all progress following that point becomes invalid. However, unlike a classical processor, a quantum system cannot checkpoint the last valid state due to the no-cloning theorem. Instead, it would need to reverse the operations to roll back to the first faulty non-Clifford, and then re-execute from that point onward. Such a speculative paradigm may be technically feasible in regions of the program without irreversible operations such as measurements or resets, which are relatively sparse in our workloads. Nevertheless, it would introduce nontrivial complexity into execution and recovery, making it an interesting direction for future work.
}

\section{Conclusion}
    Recent work has proposed \textit{Yoked Surface Codes (YSCs)} to reduce the
    physical qubit footprint of surface codes by concatenating
    the surface code with a high-rate quantum code.
    YSCs have 1D and 2D variants, which are $1.8\times$ and $4.5\times$
    more efficient than the surface code, respectively.
    As the 2D YSC has superior storage efficiency compared to the 1D YSC,
    modern \textit{Fault-Tolerant Quantum Computer (FTQC)} architectures
    have evolved to use a two-level memory hierarchy where surface codes
    perform logical operations and 2D YSCs store idle program qubits.
    Loading from the 2D YSC first requires \textit{retrieval} which
    moves a program qubit out of the 2D YSC memory and into a surface
    code qubit.
    However, retrieval is not fault-tolerant, so the FTQC must \textit{verify}
    that the program qubit was loaded correctly.
    Verification dominates memory access latency, and eliminating it entirely
    would make the FTQC $5.7\times$ faster.
    To this end, we propose Trident which is a three-level memory hierarchy
    with an intermediate 1D YSC memory tier.
    Our insight is that 1D YSCs have zero verification latency.
    We propose two versions of Trident which operate the memory hierarchy
    differently: \textit{Trident-D (Demand-Driven Hierarchy)} and
    \textit{Trident-P (Prefetch-Driven Hierarchy)}.
    Trident-D and Trident-P provide speedups of $2.8\times$ and $4.2\times$ over the
    baseline.
    Both versions of Trident only require $4.9\%$ additional physical qubits to implement.

\ignore{
    The surface code is a promising quantum error correcting code expected
    to be used in future Fault-Tolerant Quantum Computers (FTQCs).
    However, the encoding rate of the surface code is rather poor, and the
    large physical qubit footprint of the surface code is an impediment
    to realizing FTQCs at scale.
    To this end, recent work has proposed \textit{Yoked Surface Codes (YSCs)}
    which mitigate the overheads of surface codes by concatenating the
    surface code with a high-rate quantum code.
    YSCs have 1D and 2D variants, which are $1.8\times$ and $4.5\times$
    more efficient than the surface code, respectively.

    Modern FTQC architectures have evolved to use a two-level hierarchy
    where surface codes perform logical operations in the compute subsystem
    and idle program qubits are stored using 2D YSCs in the memory subsystem.
    Loading a program qubit from a 2D YSC memory first requires \textit{retrieval},
    which loads a program qubit out of the 2D YSC and into a surface code
    logical qubit.
    However, retrieval is not fault-tolerant as the 2D YSC cannot measure its
    yokes concurrently with the retrieval operation.
    Consequently, the FTQC must \textit{verify} the load to ensure the program
    qubit was loaded in the correct state.
    For 2D YSCs, load verification takes 178 cycles and is much longer than
    the 8 cycle latency of load retrieval.
    Thus, load verification dominates memory access latency.
    If we were to eliminate verification latency entirely, the FTQC would be
    $5.7\times$ faster on average.

    Our observation is that 1D YSCs, while less storage-efficient than 2D YSCs,
    have zero verification latency.
    Ideally, we desire the zero verification latency of 1D YSCs while retaining
    the efficiency of 2D YSCs.
    To this end, we propose \textit{Trident} which expands the two-level memory
    hierarchy into a three-level memory hierarchy.
    The intermediate memory tier consists of a single, 8-entry 1D YSC.
    If a majority of loads access the intermediate L1 memory rather than the 
    2D YSC (L2) memory, load verification latency will greatly decrease on average.
    We propose two variants of Trident which manage the memory hierarchy differently.
    We first propose \textit{Trident-D (Demand-Driven Trident)} which stores
    the recently accessed program qubits in the L1 memory, which operates as a
    cache for the L2.
    We also propose \textit{Trident-P (Prefetch-Driven Trident)} which instead
    operates the L1 memory as a prefetch buffer.
    Trident-D and Trident-P provide speedups of $2.8\times$ and $4.2\times$ over the
    baseline.
    Both versions of Trident only require $4.9\%$ additional physical qubits to implement.
}

\section{Acknowledgements}

K.D. and K.N.S. were supported by Google’s Research Scholar Program.

\bibliographystyle{ACM-Reference-Format}
\balance
\bibliography{refs}

\appendix

\newpage
\section{Artifact Appendix}

\subsection{Abstract}

Our artifact is Quicksilver, a software package for compiling and simulating quantum programs on architectures with separate compute and memory subsystems. Quicksilver compiles quantum programs into a binary instruction format and simulates their execution in detail. It implements the memory compilation passes for Trident-D and Trident-P and models the runtime behavior of YSC memory. We also provide the benchmark programs used in our evaluations.

The artifact supports reproduction of all quantitative results in the paper: Figures~\ref{fig:background}(c), \ref{fig:lru_opt_speedup}, \ref{fig:performance}, \ref{fig:verification}, \ref{fig:coverage}, \ref{fig:breakdown}, \ref{fig:buffer_capacity}, \ref{fig:compute_capacity},
and Tables~\ref{tab:block_size_sens}, \ref{tab:fidelity}, \ref{tab:storageeff}. We provide a single Python script to run the experiments and generate plots in an iPython notebook.

\subsection{Artifact check-list (meta-information)}

{\small
\begin{itemize}
  \item {\bf Program: } Benchmark programs are available at \href{https://doi.org/10.5281/zenodo.22931705}{10.5281/zenodo.22931705}. 
  \item {\bf Compilation: }CMake, C++23.
  \item {\bf Transformations: }A two-pass memory scheduler inserts loads and stores in the first pass, and movement between L1 and L2 in the second pass.
  \item {\bf Hardware: }A CPU host, (Optional) compute cluster.
  \item {\bf Metrics: } Program throughput, hit rate, logical error rate. 
  \item {\bf Output: }Plots and tables corresponding to the paper's quantitative results. Expected results are included.
  \item {\bf Experiments: }A Python script automates evaluation.
  \item {\bf How much disk space is required (approximately)?: }At most 50 GB.
  \item {\bf How much time is needed to prepare workflow (approximately)?: } 30 minutes.
  \item {\bf How much time is needed to complete experiments (approximately)?: } One day on a compute cluster.
  \item {\bf Publicly available?: } Yes
  \item {\bf Code licenses (if publicly available)?: } MIT
  \item {\bf Data licenses (if publicly available)?: } MIT
  \item {\bf Archived (provide DOI)?: } \href{https://doi.org/10.5281/zenodo.22933092}{10.5281/zenodo.22933092}
\end{itemize}
}

\subsection{How to access}
Clone the 
\href{https://github.com/gunagya/quicksilver/tree/asplos2027}{Quicksilver Github repository} and switch to the \texttt{asplos2027} branch.
Download the benchmark binaries from \href{https://doi.org/10.5281/zenodo.22931705}{Zenodo} and place the files in \texttt{benchmarks/bin/}.

\subsection{Dependencies}

\subsubsection{Software dependencies}

The build requires CMake 3.20.3 or newer, and a C++23 compiler (e.g. GCC 13). The experiment scripts require Python 3.10 or newer. Plot generation requires Matplotlib 3.7 and a Jupyter environment, such as JupyterLab. The Python dependencies are listed in \texttt{requirements-artifact.txt}.

\subsubsection{Data sets}

We provide compressed binaries for all benchmark programs on \href{https://doi.org/10.5281/zenodo.22931705}{Zenodo}. These binaries are generated from QASM descriptions of representative portions of quantum programs. The programs are first compiled to Clifford+$R_Z$, after which Gridsynth decomposes the $R_Z$ rotations into Clifford+$T$ sequences. The resulting binaries contain the program's \textit{computational instructions}.

\subsection{Installation and Testing}

\subsubsection{Installation}

After cloning the repository and downloading the benchmarks, run the following commands  from the repository root to install dependencies and build the executables. For more details, refer to the repository \href{https://github.com/gunagya/quicksilver/blob/asplos2027/README.md}{README}.
\begin{verbatim}
    
$ python3 -m venv venv
$ source venv/bin/activate
$ python3 -m pip install -r \
    requirements-artifact.txt
$ cmake -S . -B build \
    -DCMAKE_BUILD_TYPE=Release \
    -DCOMPILE_BINARY_GENERATION=OFF
$ cmake --build build --target \
    qs_memory_scheduler yoked_simulator -j2
\end{verbatim}

\subsubsection{Basic Testing}
To validate the installation, compile and simulate the first 100 instructions of \texttt{bose\_hubbard\_q}:

\begin{verbatim}
$ python3 run_artifact.py --instructions 100 \
    --benchmarks bose_hubbard_q --plots
\end{verbatim}

Verify that outputs are written to 
\texttt{results/smoke\_100/}.


\subsection{Evaluation}
\subsubsection{Experiment workflow}

The experiment runner executes two phases:
\begin{enumerate}
    \item \textit{Compilation}: The input binaries undergo two memory compilation passes. The first pass inserts loads and stores between the compute and memory subsystems according to the specified compute capacity. The output binary is added to \texttt{firstpass} under \texttt{benchmarks/bin}. The second pass augments outputs of \texttt{firstpass} by adding movement between L1 and L2 according to a \texttt{cache} (Trident-D) or \texttt{prefetch} (Trident-P) strategy, and writes output binaries to \texttt{secondpass}.
    \item \textit{Simulation}: The simulator models cycle-level execution of the compiled binaries using gate latencies, instruction dependencies, and resource availability. It uses the \texttt{firstpass} binaries to evaluate the baseline and idealized systems, and the \texttt{secondpass} binaries to evaluate Trident-D and Trident-P. Each simulation records its metrics in a log file under \texttt{build}.
\end{enumerate}

\subsubsection{Evaluation and expected results}
To reproduce the results of the paper, run:

\begin{verbatim}
python3 run_artifact.py --plots
\end{verbatim}

The script compiles all benchmarks and simulates the first 100 million instructions for each configuration. It sweeps over values of L1 capacity, compute capacity, and L2 block size to reproduce the sensitivity studies.

All results can be found by examining their respective cells in the iPython notebook \texttt{results/paper\_plots.ipynb}. Generated plots and data should roughly match reported results, with some possible variance due
to randomness.

\subsubsection{Experiment customization}

Compilation and simulation parameters, including those varied in the sensitivity studies, are specified through command-line arguments. To evaluate other configurations, modify the arguments in the commands printed by \texttt{run\_artifact.py}. Settings that control YSC memory behavior (e.g. maximum idle time, frequency of yoke checks) can be changed in the source code and require rebuilding the executables.

\subsection{Methodology}

Submission, reviewing and badging methodology:

\begin{itemize}
  \item \url{https://www.acm.org/publications/policies/artifact-review-badging}
  \item \url{http://cTuning.org/ae/submission-20201122.html}
  \item \url{http://cTuning.org/ae/reviewing-20201122.html}
\end{itemize}

\section{Syndrome History Criticality}
\label{sec:app:measurement_resolution}

Reliably executing a logical operation on a logical qubit requires the FTQC to have the entire syndrome history of that qubit.
If the FTQC were to execute the operation incorrectly, then a Pauli correction (i.e., $X$) must be applied to the qubit.
If this Pauli correction commutes through the next logical operation, then the FTQC can perform the current logical
operation without having resolved the syndrome history of the qubit.
Then, the syndrome history can be resolved later, and the Pauli correction can be applied upon resolution.
However, resolving the syndrome history becomes critical when the potential Pauli correction cannot be passed through
the next logical operation.
This situation arises in $RZ$ and $CCX$ gates where $T$ gates prevent $X$ corrections from passing through.
Thus, when the FTQC encounters a $T$ gate, it must stall until the syndrome history of the logical qubit is resolved.

This is relevant to load verification in YSCs as the syndrome history of the loaded program qubit is incomplete until
verification completes.
While the loaded program qubit can execute Clifford gates like $H$ or $CX$ before verification completes,
it cannot execute $T$ gates.
\textit{We model this aspect of load verification in our evaluations. In practice, program qubits quickly encounter
a $T$ gate after being loaded, so verification latency cannot be hidden.}

\ignore{
After load retrieval from a YSC block, the loaded state must eventually be verified, since the load may have introduced an error. However, the latency of this verification does not always lie on the critical path of execution. An error during the load manifests as a Pauli error on the qubit transferred into the compute region. Such an error can be tracked and corrected by updating the \emph{Pauli frame} maintained in software.

If the loaded qubit next participates in only Clifford operations, then the Pauli frame update can be safely delayed. This is because Pauli corrections can be propagated through Clifford gates without changing the underlying Clifford operations; only the resulting Pauli correction changes. So, if a later yoke check during load verification reveals that an error occurred during the load, the required Pauli correction can be pushed through any applied Clifford gates and absorbed as an update to the Pauli frame. Thus, verification latency does not stall Clifford operations on a loaded qubit.

The situation changes if the loaded qubit participates in a non-Clifford operation. In general, Pauli corrections cannot be propagated through a non-Clifford gate while preserving the gate sequence. For example, propagating an $X$ correction across a $T$ gate changes it to a $T^\dagger$. If verification is delayed past a non-Clifford and later yoke check reveals that a load error occurred, recovery would require undoing and re-executing the sequence of operations already applied to the qubit. Such recovery is undesirable and may not always be possible. Consequently, the loaded state must be fully verified before the qubit participates in a non-Clifford operation. If the source YSC block has not yet checked all its yokes following the load, the qubit must stall until verification is complete.
}

\section{2D YSC:  Operations for Load Retrieval}
\label{appendix:2d_load}

Figure~\ref{fig:2d_ysc_load} illustrates the sequence of operations required to load a logical qubit \texttt{qA} from 2D YSC into a surface code qubit \texttt{qLD}. In a 2D YSC, both the $X$ and $Z$ operators of a logical qubit are distributed across four surface code patches. 

Panels 1--3 implement the \emph{first retrieval step}. This step measures the joint $Z$ parity between \texttt{qLD} and the four surface code patches supporting the logical $Z$ operator of \texttt{qA}, highlighted in blue. Due to limited routing space in 2D YSCs, this measurement must be split across two lattice surgery operations. In panel 1, \texttt{qLD} is initialized to $|+\rangle$, and lattice surgery measures the joint $Z$ parity between an ancilla and two of the four $Z$-supporting patches of \texttt{qA}. To retain the same protection as the YSC memory, this lattice surgery must be carried out in an ancilla region with width and height equal to the effective code distance ($d_\text{eff}$) of the YSC, and its intermediate parity checks must be measured for $d_\text{eff}$ syndrome extraction rounds. Throughout this paper, we treat $d_\text{eff}$ syndrome extraction rounds as one cycle. In panel 2, the routing space is moved adjacent to the remaining two $Z$-supporting patches using sliding operations, which require $2d_\text{eff}$ rounds, or 2 cycles \cite{dubey2025_access}. Panel 3 then completes the $ZZ$ retrieval by measuring the joint $Z$ parity of \texttt{qLD} with the remaining two patches and the ancilla, which takes 1 cycle.

\begin{figure}[h]
    \centering
    \includegraphics[width=1\linewidth]{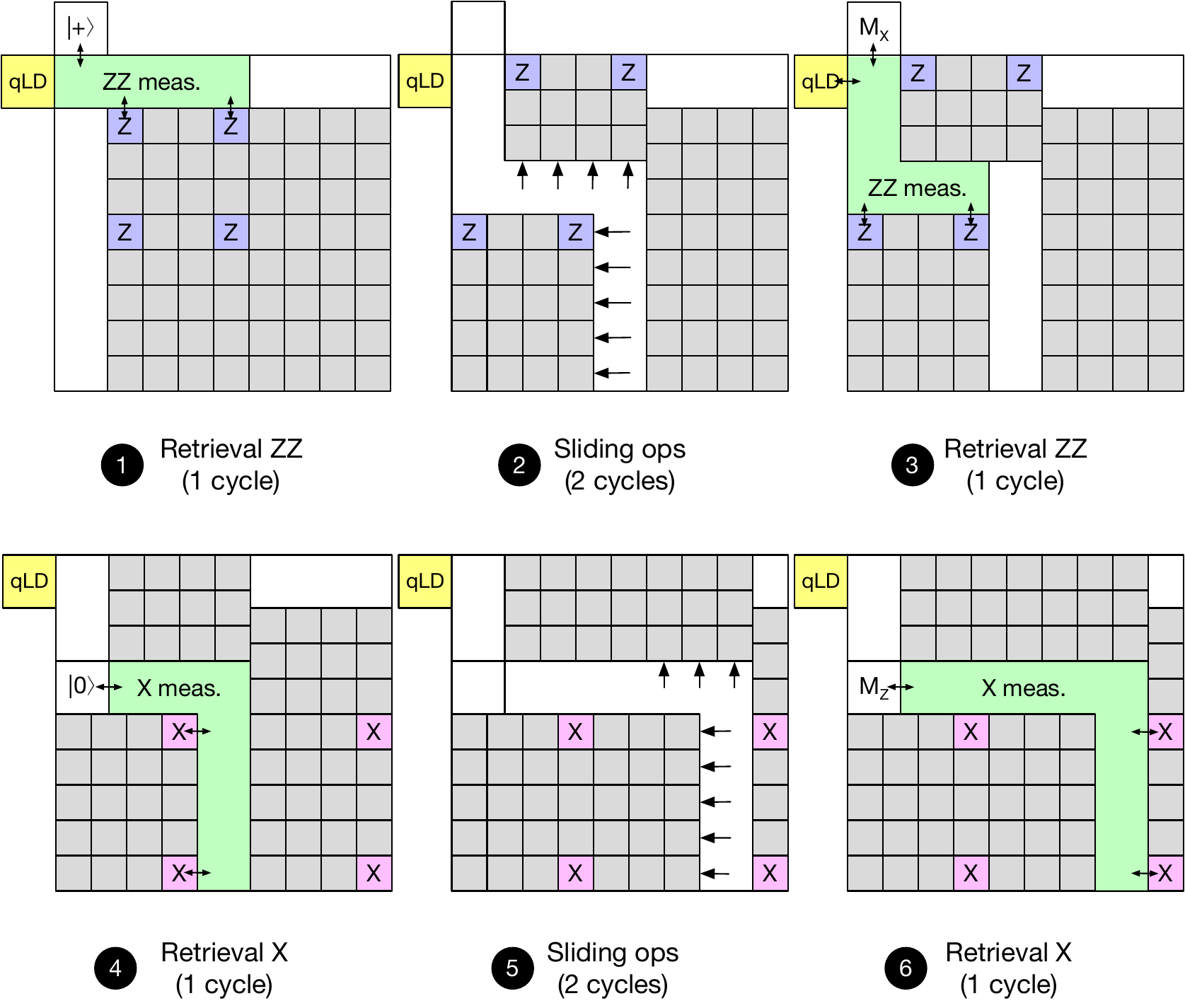}
    \caption{Operations for load retrieval from 2D YSC. The retrieval completes in eight cycles.}
    \label{fig:2d_ysc_load}
\end{figure}

Panels 4--6 implement the \emph{second retrieval step}, which measures \texttt{qA} in the $X$ basis. This is done by measuring the joint $X$ parity of the four patches supporting the logical $X$ operator of \texttt{qA}, highlighted in pink. Like before, this is done with two lattice surgery operations. An ancilla patch is initialized in $|0\rangle$ in panel 4 and is finally measured in the $Z$ basis at the end of panel 6. At that point, the state of \texttt{qA} is loaded to \texttt{qLD}. The full load retrieval sequence takes eight cycles. 

Unlike a load retrieval from 1D YSC, this retrieval does not immediately verify the loaded state. Verification requires the yokes of the 2D YSC to be measured following the retrieval.

\section{Fragmentation Study: 1D YSC Overheads}
\label{appendix:fragmentation}

In this section, we describe how 1D YSCs can reduce fragmentation inherent in 2D YSC memory, and halve the additional physical qubit cost of Trident on average. 

The discrete capacities of 2D YSC blocks can require memory overprovisioning, leading to \emph{internal fragmentation}. For example, \texttt{ethylene\_t} stores 237 qubits in memory. Using standard $[[256, 194, 4]]$ 2D YSCs requires two such blocks, although the second holds only 43 qubits. A smaller second block could save space, but the smallest block that accommodates these remaining qubits is $[[144,98,4]]$, since the block width must be a multiple of four. Thus, storing solely with 2D YSC blocks is inefficient. In contrast, 1D YSCs support any even capacity. The minimum-cost configuration therefore combines two 2D YSC blocks, $[[256,194,4]]$ and $[[64,34,4]]$, with one $[[12,10,2]]$ 1D YSC block. Here, including a 1D YSC minimizes the physical qubit footprint even without considering Trident's performance benefits. Trident can use this block as L1 at no additional physical qubit cost.

To quantify this benefit, we implement a memory configuration optimizer that minimizes the physical qubit footprint for a target memory capacity. The optimizer first enumerates all supported 1D and 2D YSC block sizes and precomputes the minimum physical qubit cost for each block capacity. It then uses dynamic programming to select the lowest-cost combination of blocks that meets or exceeds the target capacity. To determine Trident's overhead, we repeat the optimization with the constraint that the configuration include a 1D YSC block whose capacity is at least the desired L1 buffer size.

Table~\ref{tab:fragmentation} reports the optimal memory configurations and additional physical qubits required by Trident. For four of the eight benchmarks, the optimal configuration contains only 2D YSC blocks; provisioning an eight-entry L1 for Trident requires 8.8k additional physical qubits. The other four configurations already include a 1D YSC block with at least eight entries. Trident can manage this existing block as L1, incurring no additional storage overhead.

\begin{table}[H]
    \begin{center}
    \caption{Trident's physical qubit overhead relative to optimal memory configurations. Half of the configurations already include a 1D YSC with at least eight entries, eliminating the additional storage cost of L1.}
    \label{tab:fragmentation}
    \footnotesize
    \begin{tabular}{|c|c|c|}
        \hline
        \multirow{2}{*}{Benchmark} & Optimal Memory  & Trident Overhead\\
        & Configuration & (physical qubits) \\
        \hline
        \hline

        \multirow{2}{*}{bose\_hubbard\_q}
            & \textbf{2D}: $11\times [[256,194,4]],$
            & \multirow{2}{*}{+8800} \\
            & $1\times [[144,98,4]]$ & \\
        \hline

        chromium\_q
            & \textbf{2D}: $1\times [[256,194,4]]$
            & +8800 \\
        \hline

        \multirow{2}{*}{ethylene\_q}
            & \textbf{2D}: $1\times [[256,194,4]],$
            & \multirow{2}{*}{+8800} \\
            & $1\times [[144,98,4]]$ & \\
        \hline

        \multirow{3}{*}{bose\_hubbard\_t}
            & \textbf{2D}: $11\times [[256,194,4]],$
            & \multirow{3}{*}{0} \\
            & $1\times [[64,34,4]]$ & \\
            & \textbf{1D}: $[[18,16,2]]$ & \\
        \hline

        \multirow{2}{*}{chromium\_t}
            & \textbf{2D}: $1\times [[144,98,4]]$
            & \multirow{2}{*}{0} \\
            & \textbf{1D}: $[[22,20,2]]$ & \\
        \hline

        \multirow{3}{*}{ethylene\_t}
            & \textbf{2D}: $1\times [[256,194,4]],$
            & \multirow{3}{*}{0} \\
            & $1\times [[64,34,4]]$ & \\
            & \textbf{1D}: $[[12,10,2]]$ & \\
        \hline

        \multirow{2}{*}{shor}
            & \textbf{2D}: $1\times [[144,98,4]]$
            & \multirow{2}{*}{0} \\
            & \textbf{1D}: $[[12,10,2]]$ & \\
        \hline

        grover
            & \textbf{2D}: $9\times [[256,194,4]]$
            & +8800 \\
        \hline
    \end{tabular}
    \end{center}
\end{table}

To evaluate beyond our benchmark suite, we sweep target memory capacities from 100 to 2500 logical qubits. Across this range, requiring an eight-entry 1D YSC for Trident adds an average of 4.4k physical qubits, half the block's nominal cost of 8.8k physical qubits. Thus, after accounting for fragmentation, Trident requires no additional physical qubits in the best case, 4.4k on average, and 8.8k in the worst case.

\end{document}